\documentclass[5p, twocolumn, times, sort&compress]{elsarticle}
\usepackage{url}  
\usepackage{amsmath}
\usepackage{amsthm}
\usepackage{amssymb}
\usepackage{amsfonts}
\usepackage{mathtools}
\usepackage[overload]{empheq}
\usepackage[section]{placeins} 
\usepackage{tikz}  
\usepackage{relsize}  
\usepackage{makecell}  
\usepackage{stfloats}  
\usepackage{booktabs}  
\usepackage{threeparttable}  
\usepackage[group-separator={,}]{siunitx}  
\usepackage[multidot]{grffile}

\usepackage{lipsum}  

\DeclareMathOperator*{\argmin}{arg\,min}  
\RenewDocumentCommand{\vec}{m}{\mathbf{#1}}  
\NewDocumentCommand{\mat}{m}{\boldsymbol{\mathit{#1}}}  
\NewDocumentCommand{\diff}{}{\mathop{}\!\mathrm{d}}  
\NewDocumentCommand{\pdiff}{mm}{\frac{\partial #1}{\partial #2}}  

\usetikzlibrary{decorations.text, arrows.meta, bending, positioning, fit, calc}

\tikzstyle{none}=[inner sep=0mm]

\tikzstyle{title}=[font=\fontsize{6}{6}\color{black!50}\ttfamily]

\tikzstyle{input}=[
    fill=white, draw=black, shape=rectangle, minimum width=0.2in, minimum height=0.2in,
    inner sep=0.02in, align=center, line width=0.01in
]

\tikzstyle{param}=[
    fill={rgb,255: red,191; green,191; blue,191}, draw=black, shape=circle, minimum width=0.25in,
    minimum height=0.25in, inner sep=0in, align=center, line width=0.01in
]

\tikzstyle{one arrow}=[->, >=stealth, line width=0.005in]

\graphicspath{{figs/}}  
\DeclareGraphicsExtensions{.pdf,.png}  

\journal{an Elsevier journal}

\begin{document}

    \begin{frontmatter}
        \title{%
        Predictive Limitations of Physics-Informed Neural Networks in Vortex Shedding%
        }

        \author[1]{Pi-Yueh Chuang}
        \ead{pychuang@gwu.edu}
        \author[1]{Lorena A. Barba\corref{cor1}}
        \ead{labarba@gwu.edu}
        \cortext[cor1]{Corresponding author}
        \affiliation[1]{%
            organization={%
                Department of Mechanical and Aerospace Engineering, %
                The George Washington University%
            }, %
            city={Washington}, %
            state={DC 20052}, %
            country={USA}%
        }

        \begin{abstract}
            The recent surge of interest in physics-informed neural network (PINN) methods has led to a wave of studies that attest to their potential for solving partial differential equations (PDEs) and predicting the dynamics of physical systems. However, the predictive limitations of PINNs have not been thoroughly investigated. We look at the flow around a 2D cylinder and find that data-free PINNs are unable to predict vortex shedding. Data-driven PINN exhibits vortex shedding only while the training data (from a traditional CFD solver) is available, but reverts to the steady state solution when the data flow stops. We conducted dynamic mode decomposition and analyze the Koopman modes in the solutions obtained with PINNs versus a traditional fluid solver (PetIBM). The distribution of the Koopman eigenvalues on the complex plane suggests that PINN is numerically dispersive and diffusive. The PINN method  reverts to the steady solution possibly as a consequence of spectral bias. This case study reaises concerns about the ability of PINNs to predict flows with instabilities, specifically vortex shedding. Our computational study supports the need for more theoretical work to analyze the numerical properties of PINN methods. The results in this paper are transparent and reproducible, with all data and code available in public repositories and persistent archives; links are provided in the paper repository at \url{https://github.com/barbagroup/jcs_paper_pinn}, and a Reproducibility Statement within the paper.
        \end{abstract}

    \end{frontmatter}

    \section{Introduction}

In recent years, research interest in using Physics-Informed Neural Networks (PINNs) has surged.
The idea of using neural networks to represent solutions of ordinary and partial differential equations goes back to the 1990s \cite{dissanayake_neural-network-based_1994,lagaris_artificial_1998}, but upon the term PINN being coined about five years ago, the field exploded. 
Partly, it reflects the immense popularity of all things machine learning and artificial intelligence (ML/AI). 
It also seems very attractive to be able to solve differential equations without meshing the domain, and without having to discretize the equations in space and time. 
PINN methods incorporate the differential equations as constraints in the loss function, and obtain the solution by minimizing the loss function using standard ML techniques.
They are easily implemented in a few lines of code, taking advantage of the ML frameworks that have become available in recent years, such as PyTorch. 
In contrast, traditional numerical solvers for PDEs such as the Navier-Stokes equations can require years of expertise and thousands of lines of code to develop, test and maintain. 
The general optimism in this field has perhaps held back critical examinations of the limitations of PINNs, and the challenges of using them in practical applications. 
This is compounded by the well-known fact that the academic literature is biased to positive results, and negative results are rarely published. 
We agree with a recent perspective article that calls for a view of ``cautious optimism'' in these emerging methods \cite{vinuesa_emerging_2022}, for which discussion in the published literature of both successes and failures is needed.

In this paper, we examine the solution of Navier-Stokes equations using PINNs in flows with instabilities, particularly vortex shedding. 
Fluid dynamic instabilities are ubiquitous in nature and engineering applications, and any method competing with traditional CFD should be able to handle them. 
In a previous conference paper, we already reported on our observations of the limitations of PINNs in this context \cite{chuang_experience_2022}. 
Although the solution of a laminar flow with vorticity, the classical Taylor-Green vortex, was well represented by a PINN solver, the same network architecture failed to give the expected solution in a flow with vortex shedding. 
The PINN solver accurately represented the steady solution at a lower Reynolds number of $Re=40$, but reverted to the steady state solution in two-dimensional flow past a circular cylinder at $Re=200$, which is known to exhibit vortex shedding. 
Here, we investigate this failure in more detail, comparing with a traditional CFD solver and with a data-driven PINN that receives as training data the solution of the CFD solver. 
We look at various fluid diagnostics, and also use dynamic mode decomposition (DMD) to analyze the flow and help explain the difficulty of the PINN solver to capture oscillatory solutions.

Other works have called attention to possible failure modes for PINN methods. Krishnapriyan el al. \cite{krishnapriyan_failure_2021} studied PINN models of simple problems of convection, reaction, and reaction-diffusion, and found that the PINN method only works for the simplest, slowyly varying problems.
They suggested that the neural network architecture is expressive enough to represent a good solution, but the landscape of the loss function is too complex for the optimization to find it. 
Fuks and Tchelepi \cite{fuks_limitations_2020} studied the limitations of PINNs in solving the Buckley-Leverett equation, a nonlinear hyperbolic equation that models two-phase flow in porous media. 
They found that the neural network model was unable to represent the solution of the 1D hyperbolic PDE when shocks were present, and also concluded that the problem was the optimization process, or the loss function.
The failure to capture the vortex shedding of cylinder flow is also highlighted in a recent work by Rohrhofer et al. \cite{rohrhofer_fixedpoints_2023}, who cite our previous conference paper. 

Our PINN solvers were built using the NVIDIA \emph{Modulus} toolkit,\footnote{\url{https://developer.nvidia.com/modulus}} a high-level package built on PyTorch for building, training, and fine-tuning physics-informed machine learning models.
For the traditional CFD solver, we used our own code, \emph{PetIBM}, which is open-source and available on GitHub, and has also been peer reviewed \cite{chuang_petibm_2018}. 
A Reproducibility Statement gives more details regarding all the open research objects to accompany the paper, and how the interested reader can reuse them.

    \section{Method}

We will be solving the 2D incompressible Navier-Stokes equations in primitive-variable form:
\begin{empheq}[left=\left\{\,, right=\right.]{equation}\label{eq:orig-ns}
    \begin{aligned}
    &\nabla \cdot \vec{u} = 0 \\
    &\pdiff{\vec{u}}{t} + \left(\vec{u} \cdot \nabla\right) \vec{u}
        =
        -\frac{1}{\rho}\nabla p + \nu \nabla^2 \vec{u}
    \end{aligned}
\end{empheq}
\noindent Here, $\vec{u} \equiv \left[ u \enspace v \right]^\mathsf{T}$, $p$, $\nu$, and $\rho$ denote the velocity vector, pressure, kinematic viscosity, and the density, respectively.
Let $\vec{x} \equiv \left[ x \enspace y \right]^\mathsf{T} \in \Omega$ and $t \in \left[0,\enspace T\right]$ denote the spatial and temporal domains.
The velocity $\vec{u}$ and pressure $p$ are functions of $\vec{x}$ and $t$ for given fluid properties $\rho$ and $\nu$.
The solution to the Navier-Stokes equations is subject to initial conditions $\vec{u}(\vec{x}, t) = \left[ u_0(\vec{x}) \enspace v_0(\vec{x}) \right]^\mathsf{T}$ and $p(\vec{x}, t) = p_0(\vec{x})$ for $\vec{x} \in \Omega$ and $t=0$.
The Dirichlet boundary conditions are $u(\vec{x}, t) = u_D(\vec{x}, t)$ and $v(\vec{x}, t) = v_D(\vec{x}, t)$, on domain boundaries $\vec{x} \in \Gamma_{\displaystyle u_D}$ and $\Gamma_{\displaystyle v_D}$, respectively.
The Neumann boundary conditions are $\pdiff{u}{\vec{n}}(\vec{x}, t)=u_N(\vec{x}, t)$ and $\pdiff{v}{\vec{n}}=v_N(\vec{x}, t)$, defined on boundaries $\vec{x} \in \Gamma_{\displaystyle u_N}$ and $\Gamma_{\displaystyle v_N}$ correspondingly.
Note that in incompressible flow pressure is a Lagrangian multiplier to enforce the divergence-free condition and does not need boundary conditions theoretically.

\begin{figure*}
    \centering
    \normalsize
    \resizebox{\textwidth}{!}{\begin{tikzpicture}
	\node [none] (network) at (0, 0) {
		Network
		$\left[\begin{smallmatrix} \vec{u} \\ p \end{smallmatrix}\right] =
		G(\vec{x}, t; \vec{\theta})$
	};

	\node [below=1.5 of network.south west, input, anchor=north west] (nin1) {$x$};
	\node [below=0.5 of nin1, input] (nin2) {$y$};
	\node [below=0.5 of nin2, input] (nin3) {$t$};

	\node [right=0.3 of nin1, param] (nh12) {$h_2^1$};
	\node [above=0.5 of nh12, param] (nh11) {$h_1^1$};
	\node [below=0.5 of nh12, none]	(nh13) {$\vdots$};
	\node [below=0.5 of nh13, none]	(nh14) {$\vdots$};
	\node [below=0.5 of nh14, param] (nh15) {$h_{N_1}^1$};

	\node [above right=0.3 of nh12, none]	(nskip1) {$\cdots$};
	\node [right=0.3 of nh13, none]	(nskip2) {$\cdots$};
	\node [above right=0.3 of nh15, none]	(nskip3) {$\cdots$};

	\node [right=0.6 of nh12, param] (nh22) {$h_2^\ell$};
	\node [above=0.5 of nh22, param] (nh21) {$h_1^\ell$};
	\node [below=0.5 of nh22, none]	(nh23) {$\vdots$};
	\node [below=0.5 of nh23, none]	(nh24) {$\vdots$};
	\node [below=0.5 of nh24, param] (nh25) {$h_{N_\ell}^\ell$};

	\node [right=0.5 of nh22, input] (nout1) {$u$};
	\node [below=0.5 of nout1, input] (nout2) {$v$};
	\node [below=0.5 of nout2, input] (nout3) {$p$};

	\node [draw=black!50, fit={(network) (nin2) (nh15) (nh25) (nout2)}] (nnframe){};

	\node [above right=1.6 and 0.8 of nout1.north east, anchor=south west, input] (dudt) {
		$\frac{\partial u}{\partial t}$};
	\node [right=0.1 of dudt.east, anchor=west, input] (dudx) {
		$\frac{\partial u}{\partial x}$};
	\node [right=0.1 of dudx.east, anchor=west, input] (dudy) {
		$\frac{\partial u}{\partial y}$};
	\node [right=0.1 of dudy.east, anchor=west, input] (d2udx2) {
		$\frac{\partial^2 u}{\partial x^2}$};
	\node [right=0.1 of d2udx2.east, anchor=west, input] (d2udy2) {
		$\frac{\partial^2 u}{\partial y^2}$};

	\node [draw=black!50, fit={(dudt) (d2udy2)}] (dubox) {};

	\node [below right=1.2 and 0.8 of nout3.south east, anchor=north west, input] (dvdt) {
		$\frac{\partial v}{\partial t}$};
	\node [right=0.1 of dvdt.east, anchor=west, input] (dvdx) {
		$\frac{\partial v}{\partial x}$};
	\node [right=0.1 of dvdx.east, anchor=west, input] (dvdy) {
		$\frac{\partial v}{\partial y}$};
	\node [right=0.1 of dvdy.east, anchor=west, input] (d2vdx2) {
		$\frac{\partial^2 v}{\partial x^2}$};
	\node [right=0.1 of d2vdx2.east, anchor=west, input] (d2vdy2) {
		$\frac{\partial^2 v}{\partial y^2}$};

	\node [draw=black!50, fit={(dvdt) (d2vdy2)}] (dvbox) {};

	\node [above=0.5 of dvdt.north west, anchor=south west, input] (dpdx) {
		$\frac{\partial p}{\partial x}$};
	\node [right=0.1 of dpdx.east, anchor=west, input] (dpdy) {
		$\frac{\partial p}{\partial y}$};

	\node [draw=black!50, fit={(dpdx) (dpdy)}] (dpbox) {};

	\node [draw=black!50, fit={(dubox) (dvbox)}] (dervbox) {};

	\node [right=4.75 of nnframe.east, anchor=south west, input] (loss5) {$
		L_5 = v - v_0
		\text{\enspace if } t = 0
	$};

	\node [above=0.2 of loss5.north west, anchor=south west, input] (loss4) {$
		L_4 = u - u_0
		\text{\enspace if } t = 0
	$};

	\node [above=0.2 of loss4.north west, anchor=south west, input] (loss3) {$
		L_3 = 
			\frac{\partial v}{\partial t} +
			\vec{u} \cdot \nabla v +
			\frac{1}{\rho}\frac{\partial p}{\partial y} -
			\nu \nabla^2 v
		\text{\enspace if } \vec{x} \in {\Omega}
	$};

	\node [above=0.2 of loss3.north west, anchor=south west, input] (loss2) {$
		L_2 = 
			\frac{\partial u}{\partial t} +
			\vec{u} \cdot \nabla u +
			\frac{1}{\rho}\frac{\partial p}{\partial x} -
			\nu \nabla^2 u
		\text{\enspace if } \vec{x} \in {\Omega}
	$};

	\node [above=0.2 of loss2.north west, anchor=south west, input] (loss1) {$
		L_1 = \nabla \cdot \vec{u} \text{\enspace if } \vec{x} \in {\Omega}
	$};

	\node [below=0.2 of loss5.south west, anchor=north west, input] (loss6) {$
		L_6 = p - p_0
		\text{\enspace if } t = 0
	$};

	\node [below=0.2 of loss6.south west, anchor=north west, input] (loss7) {$
		L_7 = u - u_D \text{\enspace if } \vec{x}\in\Gamma_{\displaystyle u_D}
	$};

	\node [below=0.2 of loss7.south west, anchor=north west, input] (loss8) {$
		L_8 = v - v_D \text{\enspace if } \vec{x}\in\Gamma_{\displaystyle v_D}
	$};

	\node [below=0.2 of loss8.south west, anchor=north west, input] (loss9) {$
		L_9 = \frac{\partial u}{\partial \vec{n}} - u_N
		\text{\enspace if } \vec{x}\in\Gamma_{\displaystyle u_N}
	$};

	\node [below=0.2 of loss9.south west, anchor=north west, input] (loss10) {$
		L_{10} = \frac{\partial v}{\partial \vec{n}} - v_N
		\text{\enspace if } \vec{x}\in\Gamma_{\displaystyle v_N}
	$};

	\node [draw=black!50, fit={(loss1) (loss2) (loss3) (loss10)}] (lossframe){};

	\node [right=0.5 of lossframe.east, anchor=west, input] (argmin) {$
		\argmin\limits_{\theta \in \Theta}
		\sum\limits_{\substack{\vec{x} \in \Omega \cup \Gamma \\ t \in T}}
		\sum\limits_{j=1}^{10} L_j^2
	$};
	\node [above=0.1 of argmin.north, anchor=south, none] (argmintxt) {Optimizing/training};

	\draw [style=one arrow] (nin1) to (nh11);
	\draw [style=one arrow] (nin1) to (nh12);
	\draw [style=one arrow] (nin1) to (nh15);
	\draw [style=one arrow] (nin2) to (nh11);
	\draw [style=one arrow] (nin2) to (nh12);
	\draw [style=one arrow] (nin2) to (nh15);
	\draw [style=one arrow] (nin3) to (nh11);
	\draw [style=one arrow] (nin3) to (nh12);
	\draw [style=one arrow] (nin3) to (nh15);

	\draw [style=one arrow] (nh21) to (nout1);
	\draw [style=one arrow] (nh21) to (nout2);
	\draw [style=one arrow] (nh21) to (nout3);
	\draw [style=one arrow] (nh22) to (nout1);
	\draw [style=one arrow] (nh22) to (nout2);
	\draw [style=one arrow] (nh22) to (nout3);
	\draw [style=one arrow] (nh25) to (nout1);
	\draw [style=one arrow] (nh25) to (nout2);
	\draw [style=one arrow] (nh25) to (nout3);

	\draw [style=one arrow] (nout1.east) to (dubox.west);
	\draw [style=one arrow] (nout2.east) to (dvbox.west);
	\draw [style=one arrow] (nout3.east) to (dpbox.west);

	\draw [style=one arrow] (dubox.east) to (loss1.west);
	\draw [style=one arrow] (dvbox.east) to (loss1.west);

	\draw [style=one arrow] (nout1.east) to (loss2.west);
	\draw [style=one arrow] (nout2.east) to (loss2.west);
	\draw [style=one arrow] (dubox.east) to (loss2.west);
	\draw [style=one arrow] (dvbox.east) to (loss2.west);
	\draw [style=one arrow] (dpbox.east) to (loss2.west);

	\draw [style=one arrow] (nout1.east) to (loss3.west);
	\draw [style=one arrow] (nout2.east) to (loss3.west);
	\draw [style=one arrow] (dubox.east) to (loss3.west);
	\draw [style=one arrow] (dvbox.east) to (loss3.west);
	\draw [style=one arrow] (dpbox.east) to (loss3.west);

	\draw[style=one arrow] (nout1.east) to (loss4.west);

	\draw[style=one arrow] (nout2.east) to (loss5.west);

	\draw[style=one arrow] (nout3.east) to (loss6.west);

	\draw [style=one arrow] (nout1.east) to (loss7.west);

	\draw [style=one arrow] (nout2.east) to (loss8.west);

	\draw [style=one arrow] (dubox.east) to (loss9.west);

	\draw [style=one arrow] (dvbox.east) to (loss10.west);

	\draw [style=one arrow] (lossframe.east) to (argmin.west);

	\node [above right=0.8 and 0.3 of nnframe.north, anchor=west, none] (adtxt) {
		Automatic differentiation
	};
	\draw [-, draw=black!50] (nnframe.north) |- (adtxt.west);
	\draw [-{Latex[length=4]}, draw=black!50] (adtxt.east) -| (dervbox.north);

\end{tikzpicture}
\unskip}
    \caption{
        A graphical illustration of the workflow in PINNs:
        $\vec{x} \equiv \left[ x \enspace y \right]^\mathsf{T} \in \Omega$ and $t \in \left[0,\enspace T\right]$ denote the spatial and temporal domains.
        $\vec{u} \equiv \left[ u \enspace v \right]^\mathsf{T}$, $p$, $\nu$, and $\rho$ represent the velocity vector, pressure, kinematic viscosity, and the density, respectively.
        $G(\vec{x}, t; \theta)$ is a neural network model that approximates the solution to the Navier-Stokes equations with a set of free model parameters denoted by $\theta$.
        The $\left\{h_1^1, \cdots, h_{N_1}^1, \cdots, h_1^\ell, \cdots, h_{N_\ell}^\ell\right\}$, called hidden layers in neural networks, can be deemed as some intermediate values or temporary results during the calculations of the approximate solutions.
        Given spatial-temporal coordinates $(x, y, t)$, the neural network returns an approximate solution $(u, v, p)$ at this coordinate.
        We then apply automatic differentiation to obtain required derivatives.
        With the approximate solutions and the derivatives, we are able to calculate the residuals (also called losses, denoted by symbol $L$) between the approximates and PDEs, as well as the initial and boundary conditions. 
        Using the aggregated squared losses, we can determine the free model parameters $\theta$ by a least-square method.
    }
    \label{fig:pinn-workflow}
\end{figure*}
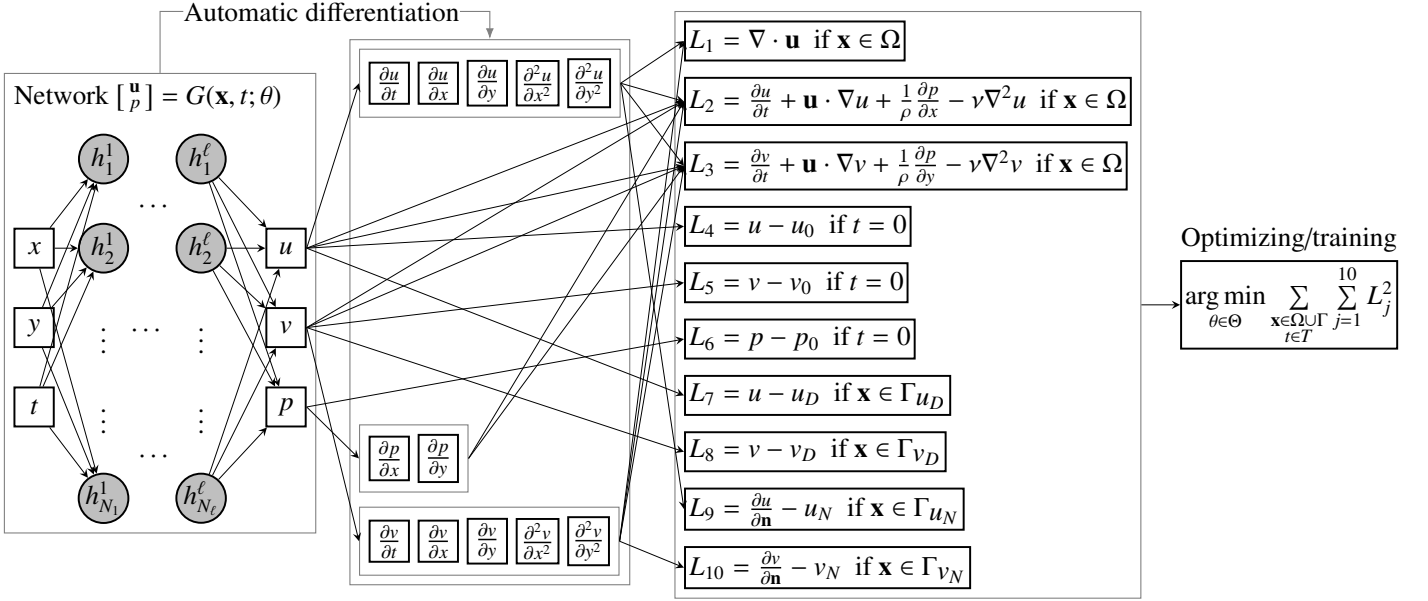

When using physics-informed neural networks, PINNs, we approximate the solutions to equation \eqref{eq:orig-ns} with a neural network model $G(\vec{x}, t; \theta)$:
\begin{equation}\label{eq:G-network}
    \begin{bmatrix}
        u(\vec{x}, t) \\ v(\vec{x}, t) \\ p(\vec{x}, t)
    \end{bmatrix}
    \approx
    G(\vec{x}, t; \theta),
\end{equation}
where $\theta$ represents a set of free model parameters we need to determine later.
A common choice of $G$ is an MLP (multilayer perceptron) network, which can be represented as follows:

\begin{gather}\label{eq:mlp-formula}
    \vec{h}^0 \equiv \begin{bmatrix} x & y & t \end{bmatrix}^\mathsf{T} \\
    \vec{h}^k =
        \sigma_{k-1}\left(\mat{A}^{k-1}\vec{h}^{k-1}+\vec{b}^{k-1}\right)
        \text{, for } 1 \le k \le \ell \\
    \begin{bmatrix} u & v & p \end{bmatrix}^\mathsf{T}
        \approx
        \vec{h}^{\ell+1} = \sigma_\ell\left(\mat{A}^\ell\vec{h}^\ell+\vec{b}^\ell\right)
\end{gather}
The vectors $\vec{h}^k$ for $1 \le k \le \ell$ are called hidden layers, and carry intermediate evaluations of the transformations that take the input (spatial and temporal variables) to the output (velocity and pressure values).
$\ell$ denotes the number of hidden layers.
The elements in these vectors are called neurons, and $N_k$ for $1 \le k \le \ell$ represents the number of neurons in each hidden layer.
To have a consistent notation, we use $\vec{h}^0$ to denote the vector of the inputs to the model $G$, which contains spatial-temporal coordinates.
Similarly, $\vec{h}^{\ell+1}$ denotes the outputs of $G$, corresponding to the approximate solutions $u$, $v$, and $p$ at every spatial point and time instant. 
$\mat{A}^k$ and $\vec{b}^k$ for $0 \le k \le \ell$ are parameter matrices and vectors holding the free model parameters that will be found via optimization, 
$\theta = \left\{ \mat{A}^0, \vec{b}^0, \cdots, \mat{A}^\ell, \vec{b}^\ell\right\}$.
Finally, $\sigma_k$ for $0 \le k \le \ell$ are vector-valued functions, called activation functions, that are applied element-wise to the vectors $\vec{h}^k$.
In neural networks, the activation functions are responsible for providing the non-linearity in an MLP model.
Throughout this work, we use $\sigma_0 = \cdots = \sigma_\ell = \sigma(\vec{z}) = \frac{\vec{z}}{1 + \exp(\vec{z})}$, the classical sigmoid function.
The parameters $\ell$, $N_k$, and the choices of $\sigma_k$ together control the model complexity of the PINNs that use MLP networks.

As with all other numerical methods for PDEs, the calculations of spatial and temporal derivatives of velocity and pressure play a crucial role.
While a numerical approximation (e.g., finite difference) may be a more robust choice---as seen in early-day literature on neural networks for differential equations \cite{dissanayake_neural-network-based_1994,lagaris_artificial_1998}---, it is common to see the use of automatic differentiation nowadays.
Automatic differentiation is a general technique to find derivatives of a function by decomposing it into elementary functions with a known derivative, and then applying the chain rule of calculus to get exact derivative of the more complex function.
Note that the word {\it exact} here refers to being exact in terms of the model $G$, rather than to the true solution of the Navier-Stokes equations. 
A detailed review of automatic differentiation can be found in reference \cite{griewank_automatic_1988}.
Major deep learning programming libraries, such as TensorFlow and PyTorch, offer the user automatic differentiation features.

Once we have obtained derivatives, we are able to calculate residuals, also called losses in the terminology of machine learning.
As shown in figure \ref{fig:pinn-workflow}, given a spatial-temporal coordinate $(x, y, t)$, we can calculate up to \num{10} loss terms, depending on where in the domain this spatial-temporal point is located. 
Figure \ref{fig:pinn-workflow} is only shown as an illustration of the PINN methodology using the solution workflow specifically for the Navier-Stokes equations \eqref{eq:orig-ns}.
The number and definitions of loss terms may change, for example, when we have some boundary segments with Robin conditions or when we are solving 3D problems.

Finally, we determine the free model parameters using a least-squares optimization, as shown in the last block in figure \ref{fig:pinn-workflow}.
To be specific, in this work we used the Adam stochastic optimization method for this process. 
We first randomly sampled some spatial-temporal collocation points from the computational domain, including all boundaries.
These points are called {\it training points} in the terminology of machine learning.
Depending on where a training point is located in the domain, it may result in multiple loss terms, as described in the previous paragraph.
An aggregated squared loss is obtained over all loss terms of all training points.
In this work, all loss terms were taken to have the same weights.
The Adam optimization then finds the optimal model parameters, i.e., $\theta=\left\{\mat{A}^0, \vec{b}^0, \cdots, \mat{A}^\ell, \vec{b}^\ell\right\}$, based on the gradients of the aggregated loss with respect to model parameters.
In other words, the desired model parameters are those giving the minimal aggregated squared loss.

Note that in figure \ref{fig:pinn-workflow} we consider that if-conditions determine the loss terms to calculate on a training point.
In practice, however, we sample points in subgroups separately from within the domain, on the boundaries, and at $t=0$.
Each subgroup of training points is only responsible for specific loss terms.
We also use a batched approach for the optimization,
meaning that not all training points are used during each individual optimization iteration.
The batched approach only uses a sample of the training points to calculate the losses and the gradients of the aggregated loss in each optimization iteration.
Hereafter, the term {\it training} will be used interchangeably with the optimization process.

In this section, we only introduce the specific details of PINNs required for our work.
References \cite{dissanayake_neural-network-based_1994,lagaris_artificial_1998,cai_physics-informed_2021} provide more details of these methods in general.


    \section{Verification and Validation}

This section presents the verification and validation (V\&V) of our PINN solvers and PetIBM, an in-house CFD solver \cite{chuang_petibm_2018}.
V\&V results are necessary to build confidence in our case study described later in section \ref{sec:case-study}.
For verification, we solved a 2D Taylor-Green vortex (TGV) at Reynolds number $Re=\num{100}$, which has a known analytical solution.
For validation, on the other hand, we use 2D cylinder flow at $Re=40$, which exhibits a well-known steady state solution with plenty of experimental data available in the published literature.

\subsection{Verification: 2D Taylor-Green Vortex (TGV), $Re=\num{100}$}\label{sec:verification}

Two-dimensional Taylor-Green vortices with periodic boundary conditions have closed-form analytical solutions,
and serve as standard verification cases for CFD solvers.
We used the following 2D TGV configuration, wih $Re=\num{100}$, to verify both the PINN solvers and PetIBM:
\begin{equation}\label{eq:tgv}
    \left\{
        \begin{aligned}
            u(x, y, t) &= \cos(x)\sin(y)\exp(-2 \nu t) \\
            v(x, y, t) &= - \sin(x)\cos(y)\exp(-2 \nu t) \\
            p(x, y, t) &= -\frac{\rho}{4}\left(\cos(2x) + \cos(2y)\right)\exp(-4 \nu t)
        \end{aligned}
    \right.
\end{equation}
where $\nu=\num{0.01}$ and $\rho=\num{1}$ are the kinematic viscosity and the density, respectively.
The spatial and temporal domains are $x, y \in \left[-\pi, \pi\right]$ and $t \in [0, 100]$.
Periodic conditions are applied at all boundaries.

In PetIBM, we used the Adams-Bashforth and the Crank-Nicolson schemes for the temporal discretization of convection and diffusion terms, respectively.
The spatial discretization is central difference for all terms.
Theoretically, we expect to see second-order convergence in both time and space for this 2D TGV problem in PetIBM.
We used the following $L_2$ spatial-temporal error to examine the convergence:
\begin{equation}\label{eq:spt-err-def}
    \begin{aligned}
    L_{2,sp-t} \equiv &\sqrt{
        \frac{1}{L_x L_y T}
        \iiint\limits_{x, y, t} \lVert f - f_{ref} \rVert^2 \diff x \diff y \diff t
    } \\
    = &
    \sqrt{\frac{1}{N_x N_y N_t}\sum\limits_{i=1}^{N_x}\sum\limits_{j=1}^{N_y}\sum\limits_{k=1}^{N_t}\left(f^{(i, j, k)} - f_{ref}^{(i, j, k)}\right)^2}
    \end{aligned}
\end{equation}
Here, $N_x$, $N_y$, and $N_t$ represent the number of solution points in $x$, $y$, and $t$;
$L_x$ and $L_y$ are the domain lengths in $x$ and $y$;
$T$ is the total simulation time;
$f$ is the flow quantity of interest, while $f_{ref}$ is the corresponding analytical solution.
The superscript $(i, j, k)$ denotes the value at the $(i, j, k)$ solution point in the discretized spatial-temporal space.
We used Cartesian grids with $2^{n} \times 2^{n}$ cells for $i=4$, $5$, $\dots$, $10$.
The time step size $\Delta t$ does not follow a fixed refinement ratio, and takes the values $\Delta t = \num{1.25e-1}$, $\num{8e-2}$, $\num{4e-2}$, $\num{2e-2}$, $\num{1e-2}$, $\num{5e-3}$, and $\num{1.25e-3}$, respectively.
$\Delta t$ was determined based on the maximum allowed CFL number and whether it was a factor of $2$ to output transient results every $\num{2}$ simulation seconds.
The velocity and pressure linear systems were both solved with BiCGSTAB (bi-conjugate gradient stabilized method).
The preconditioners of the two systems are the block Jacobi preconditioner and the algebraic multigrid preconditioner from NIVIDA's AmgX library.
At each time step, both linear solvers stop when the preconditioned residual reaches \num{e-14}.
The hardware used for PetIBM simulations contains 5 physical cores of Intel E5-2698 v4 and 1 NVIDIA V100 GPU.

\begin{figure}
    \centering%
    \includegraphics[width=\columnwidth]{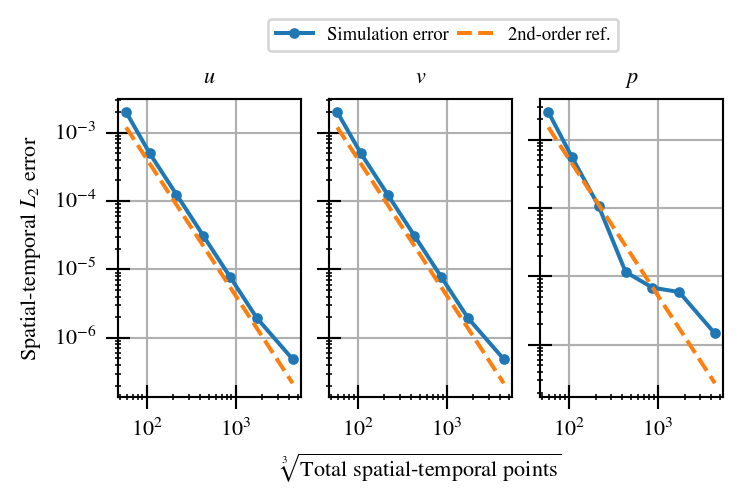}%
    \caption{%
        Grid-convergence test and verification of PetIBM using 2D TGV at $Re=\num{100}$.
        The spatial-temporal $L_2$ error is defined in equation \eqref{eq:spt-err-def}.
        Taking the cubic root of the total spatial-temporal solution points gives the characteristic cell size.
        Both $u$ and $v$ velocities follow second-order convergence, while the pressure $p$ follows the trend with some fluctuation.
    }
    \label{fig:tgv-petibm-convergence}%
\end{figure}

Figure \ref{fig:tgv-petibm-convergence} shows the spatial-temporal convergence results of PetIBM.
Both $u$ and $v$ follow an expected second-order convergence before the machine round-off errors start to dominate on the $1024 \times 1024$ grid.
The pressure follows the expected convergence rate with some fluctuations.
Further scrutiny revealed that the AmgX library was not solving the pressure system to the desired tolerance.
The AmgX library has a hard-coded stop mechanism when the relative residual (relative to the initial residual) reaches machine precision.
So while we configured the absolute tolerance to be \num{e-14}, the final preconditioned residuals of the pressure systems did not match this value.
On the other hand, the velocity systems were solved to the desired tolerance.
With this minor caveat, we consider the verification of PetIBM to be successful, as the minor issue in the convergence of pressure is irrelevant to the code implementation in PetIBM.

\begin{figure}
    \centering%
    \includegraphics[width=\columnwidth]{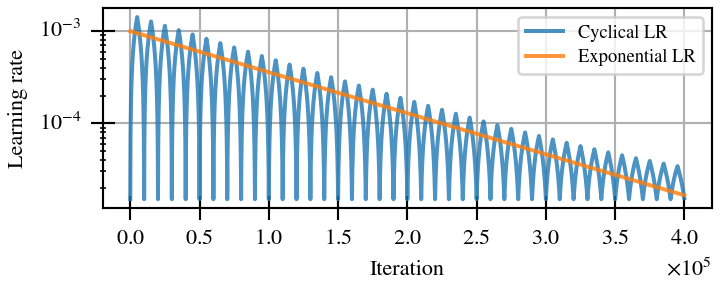}%
    \caption{%
        Learning-rate history of 2D TGV $Re=\num{100}$ w/ PINN
        The exponential learning rate scheduler is denoted as {\it Exponential}, and the cyclical scheduler is denoted as {\it Cyclical}.
    }
    \label{fig:tgv-learning-rate-hist}%
\end{figure}

Next, we solved this same TGV problem using an unsteady PINN solver.
For the optimization, we used PyTorch's Adam optimizer
with the following parameters: $\beta_1=\num{0.9}$, $\beta_2=\num{0.999}$, and $\epsilon=\num{e-8}$.
The total iteration number in the optimization is \num{400000}.
Two learning-rate schedulers were tested: the exponential learning rate and the cyclical learning rate.
Both learning rates are from PyTorch and were used merely to satisfy our curiosity.
The exponential scheduler has only one parameter in PyTorch: $\gamma=0.95^{\frac{1}{5000}}$.
The cyclical scheduler has the following parameters: $\eta_{low}=\num{1.5e-5}$, $\eta_{high}=\num{1.5e-3}$, $N_c=\num{5000}$, and $\gamma=\num{9.99989e-1}$.
These values were chosen so that the peak learning rate at each cycle is slightly higher than the exponential rates.
Figure \ref{fig:tgv-learning-rate-hist} shows a comparison of the two schedulers.

The MLP network used consisted of \num{3} hidden layers and \num{128} neurons per layer.
\num{8192e4} spatial-temporal points were used to evaluate the PDE losses (i.e., the $L_1$, $L_2$, and $L_3$ in figure \ref{fig:pinn-workflow}).
We randomly sampled these spatial-temporal points from the spatial-temporal domain$\left[-\pi, \pi\right] \times \left[-\pi, \pi\right] \times \left(0, 100\right]$.
During each optimization iteration, however, we only used \num{8192} points to evaluate the PDE losses.
It means the optimizer sees each point \num{40} times on average because we have a total of \num{4e5} iterations.
Similarly, \num{8192e4} spatial-temporal points were sampled from $x,y \in \left[-\pi, \pi\right] ] \times \left[-\pi, \pi\right]$ and $t=0$ for the initial condition loss (i.e., $L_4$ to $L_6$).
And the same number of points were sampled from each domain boundary ($x=\pm\pi$ and $y=\pm\pi$) and $t\in\left(0, 100\right]$ for boundary-condition losses ($L_7$ to $L_{10}$).
A total of \num{8192} points were used in each iteration for these losses as well.

We used one NVIDIA V100 GPU to run the unsteady PINN solver for the TGV problem.
Note that the PINN solver used single-precision floats, which is the default in PyTorch.

\begin{figure}
    \centering%
    \includegraphics[width=\columnwidth]{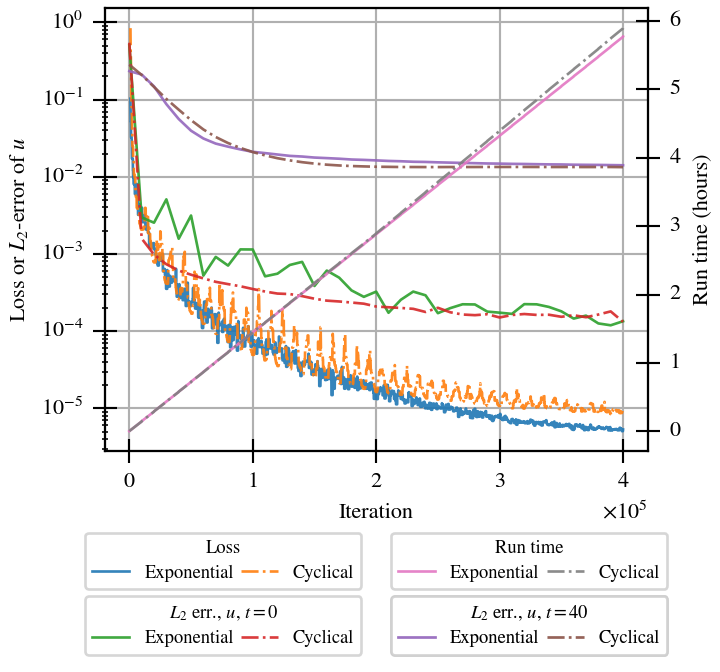}%
    \caption{%
        Histories with respect to optimization iterations for the total loss and the $L_2$ errors of $u$ at $t=0$ and $40$ in the TGV verification of the unsteady PINN solver.
        The left vertical axis corresponds to the total loss and the errors.
        The right vertical axis corresponds to the run time.
        The cyclical scheduler has a slightly better accuracy at $t=40$ with a slightly more time cost, though its total loss is higher.
    }
    \label{fig:tgv-pinn-loss}%
\end{figure}

After training, we evaluated the PINN solver's errors at cell centers in a $512$ $\times$ $512$ Cartesian grid and at $t=0$, $2$, $4$, $\cdots$, $100$.
Figure \ref{fig:tgv-pinn-loss} shows the histories of the optimization loss and the $L_2$ errors at $t=0$ and $t=40$ of the $u$ velocity on the left vertical axis.
The same figure also shows the run time (wall time) on the right vertical axis.
The total loss converges to an order of magnitude of \num{e-6}, which may reflect the fact that PyTorch uses single-precision floats.
The errors at $t=0$ and $t=40$ converge to the orders of \num{e-4} and \num{e-2}, respectively.
This observation is reasonable because the net errors over the whole temporal domain is, by definition, similar to the square root of the total, which is \num{e-3}.
The PINN solver got exact initial conditions for training (i.e., $L_4$ to $L_6$), so it is reasonable to see a better prediction accuracy at $t=0$ than later times.
Finally, though the computational performance is not the focus of this paper, for the interested reader's benefit we would like to point out that the PINN solver took about 6 hours to converge with a V100 GPU, while PetIBM needed less than 10 seconds to get an error level of \num{e-2} using a very dated K40 GPU (and most of the time was overhead to initialize the solver).

\begin{figure}[t]
    \centering%
    \includegraphics[width=0.95\columnwidth]{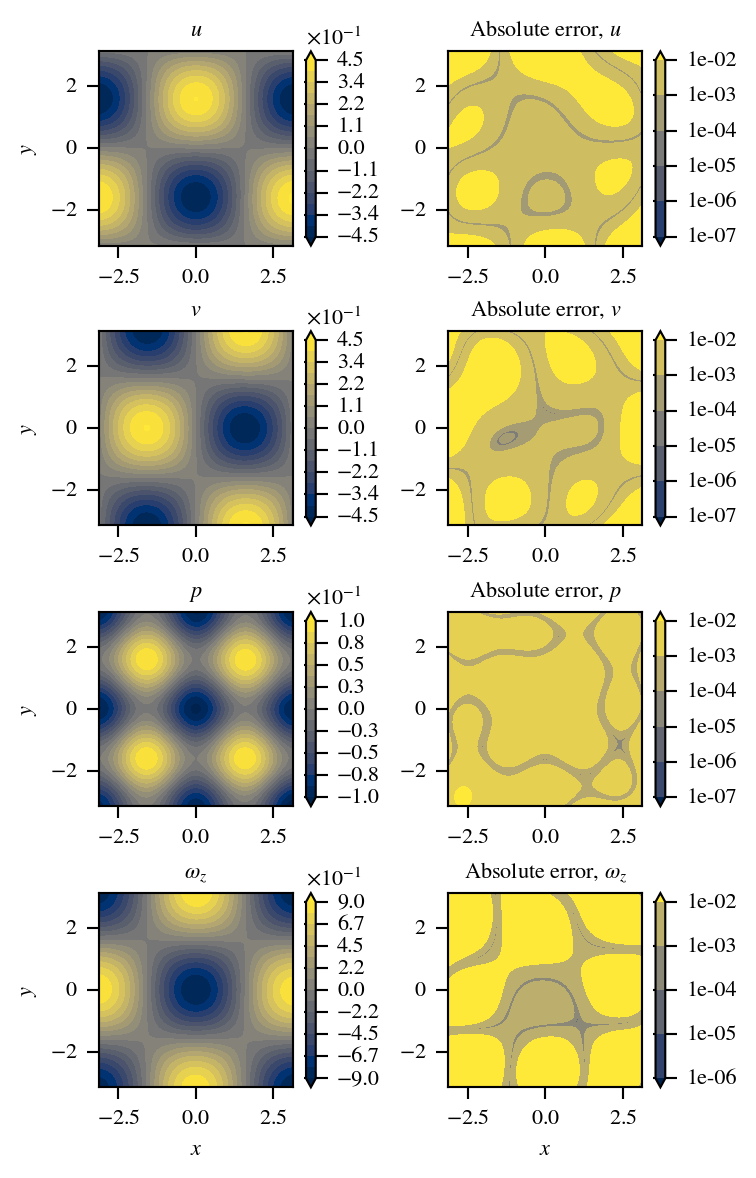}
    \caption{%
        Contours at $t=40$ of 2D TGV at $Re=\num{100}$ primitive variables and errors using the unsteady PINN solver.
        Roughly speaking, the absolute errors are at the level of \num{e-3} for primitive variables ($u$, $v$, and $p$), which corresponds to the square root of the total loss.
        The vorticity was obtained from post-processing and hence was augmented in terms of errors.
    }
    \label{fig:tgv-pinn-contours}%
\end{figure}

In sum, we determined the PINN solution to be verified, although the accuracy and the computational cost were not satisfying.
The relatively low accuracy is likely a consequence of the use of single-precision floats and the intrinsic properties of PINNs, rather than implementation errors.
Figure \ref{fig:tgv-pinn-contours} shows the contours of the PINN solver's predictions at $t=40$ for reference.

\subsection{Validation: 2D Cylinder, $Re=\num{40}$}\label{sec:val_2d_cylinder_re40}

We used 2D cylinder flow at $Re=40$ to validate the solvers because it has a similar configuration with the $Re=200$ case that we will study later.
The $Re=40$ flow, however, does not exhibit vortex shedding and reaches a steady-state solution, making it suitable for validating the core functionality of the code.
Experimental data for this flow configuration is also widely available.

The spatial and temporal computational domains are $[-10$, $30]$ $\times$ $[-10$, $10]$ and $t \in [0, 20]$.
A cylinder with a nondimensional radius of $0.5$ sits at $x=y=0$.
Density is $\rho=1$, and kinematic viscosity is $\nu=0.025$.
The initial conditions are $u=1$ and $v=0$ everywhere in the spatial domain.
The boundary conditions are $u=1$ and $v=0$ on $x=-10$ and $y=\pm 10$.
At the outlet, i.e., $x=30$, the boundary conditions are set to 1D convective conditions:
\begin{equation}\label{eq:convec-bc}
    \pdiff{}{t}\begin{bmatrix} u \\ v \end{bmatrix}
    +
    c\pdiff{}{\vec{n}}\begin{bmatrix} u \\ v \end{bmatrix} = 0,
\end{equation}
where $\vec{n}$ is the normal vector of the boundary (pointing outward), and $c=1$ is the convection speed.

We ran the PetIBM validation case on a workstation with one (very old) NVIDIA K40 GPU and 6 CPU cores of the Intel i7-5930K processor.
The grid resolution is $562 \times 447$ with $\Delta t=\num{e-2}$.
The tolerance for all linear solvers in PetIBM was $\num{e-14}$.
We used the same linear solver configurations as those in the TGV verification case.

We validated two implementations of the PINN method with this cylinder flow because both codes were used in the $Re=200$ case (section \ref{sec:case-study}).
The first implementation is an unsteady PINN solver, which is the same piece of code used in the verification case (section \ref{sec:verification}).
It solves the unsteady Navier-Stokes equations as shown in figure \ref{fig:pinn-workflow}.
The second one is a steady PINN solver, which solves the steady Navier-Stokes equations.
The workflow of the steady PINN solver works similar to that in figure \ref{fig:pinn-workflow} except that all time-related terms and losses are dropped.

Both PINN solvers used MLP networks with \num{6} hidden layers and \num{512} neurons each.
The Adam optimizer configuration is the same as that in section \ref{sec:verification}.
The learning rate scheduler is a cyclical learning rate with $\eta_{low}=\num{e-6}$, $\eta_{high}=\num{e-2}$, $N_c=\num{5000}$, and $\gamma={9.9998e-1}$.
We ran all PINN-related validations with one NVIDIA A100 GPU,
all using single-precision floats.

To evaluate PDE losses, \num{256000000} spatial-temporal points were randomly sampled from the computational domain and the desired simulation time range.
The PDE losses were evaluated on \num{25600} points in each iteration, so the Adam optimizer would see each point \num{40} times on average during the \num{400000}-iteration optimization.
On the boundaries, \num{25600000} points were sampled  at $y=\pm 10$, and \num{12800000} at $x=-10$ and $x=30$.
On the cylinder surface, the number of spatial-temporal points were \num{5120000}.
In each iteration, \num{2560}, \num{1280}, and \num{512} points were used, respectively.

\begin{figure}
    \centering%
    \includegraphics[width=\columnwidth]{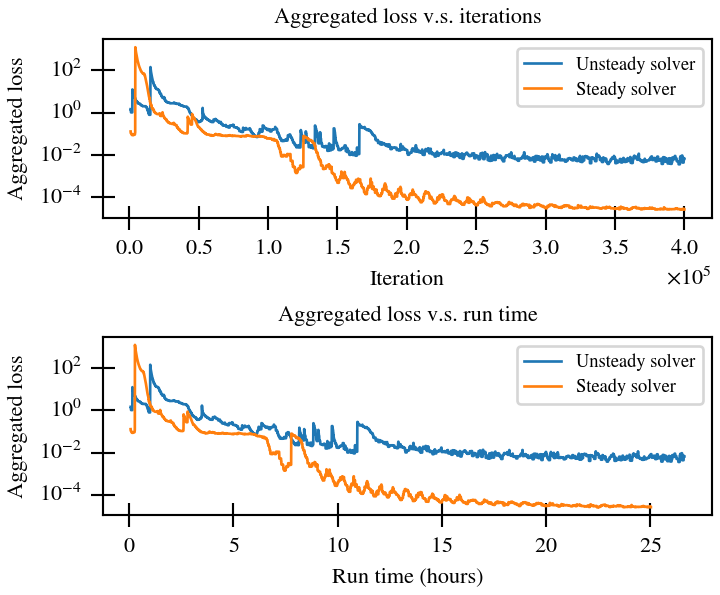}%
    \caption{%
        Training convergence history of 2D cylinder flow at $Re=\num{40}$ for both steady and unsteady PINN solvers.
    }
    \label{fig:cylinder-re40-pinn-loss}%
\end{figure}

Figure \ref{fig:cylinder-re40-pinn-loss} shows the training history of the PINN solvers.
The total loss of the steady PINN solver converged to around \num{e-4}, while that of the unsteady PINN solver converged to around \num{e-2} after about 26 hours of training.
Readers should be aware that the configuration of the PINN solvers might not be optimal, so the accuracy and the computational cost shown in this figure should not be treated as an indication of PINNs' general performance.
In our experience, it is possible to reduce the run time in half but obtain the same level of accuracy by adjusting the number of spatial-temporal points used per iteration.

\begin{figure}
    \centering%
    \includegraphics[width=\columnwidth]{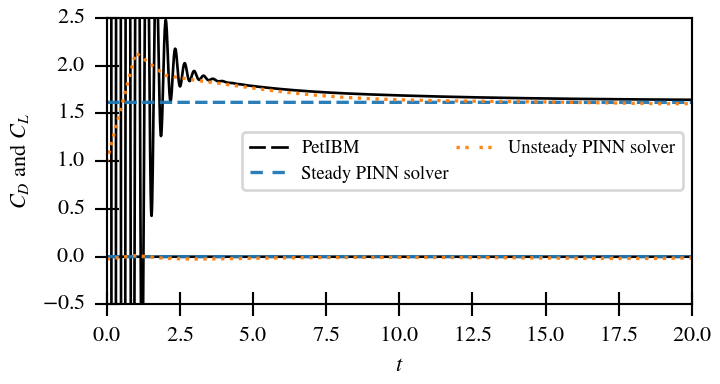}%
    \caption{%
        Drag and lift coefficients of 2D cylinder flow at $Re=\num{40}$ w/ PINNs.
    }
    \label{fig:cylinder-re40-drag-lift}%
\end{figure}

Figure \ref{fig:cylinder-re40-drag-lift} gives the drag and lift coefficients ($C_D$ and $C_L$) with respect to simulation time, where PINN and PetIBM visually agree.
Table \ref{table:cylinder-re40-cd-comparison} compares the values of $C_D$ against experimental data and simulation data from the literature.
Values from different works in the literature do not closely agree with each other.
Though there is not a single value to compare against, at least the $C_D$ from the PINN solvers and PetIBM fall into the range of other published works.
We consider the results of $C_D$ validated for the PINN solvers and PetIBM.

\begin{figure}
    \centering%
    \includegraphics[width=0.95\columnwidth]{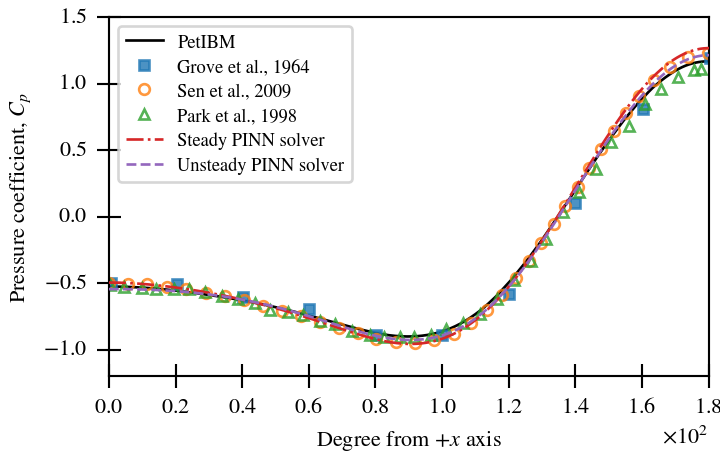}%
    \caption{%
        Surface pressure distribution of 2D cylinder flow at $Re=\num{40}$
    }
    \label{fig:cylinder-re40-pinn-surfp}%
\end{figure}

\begin{figure*}
    \centering%
    \includegraphics{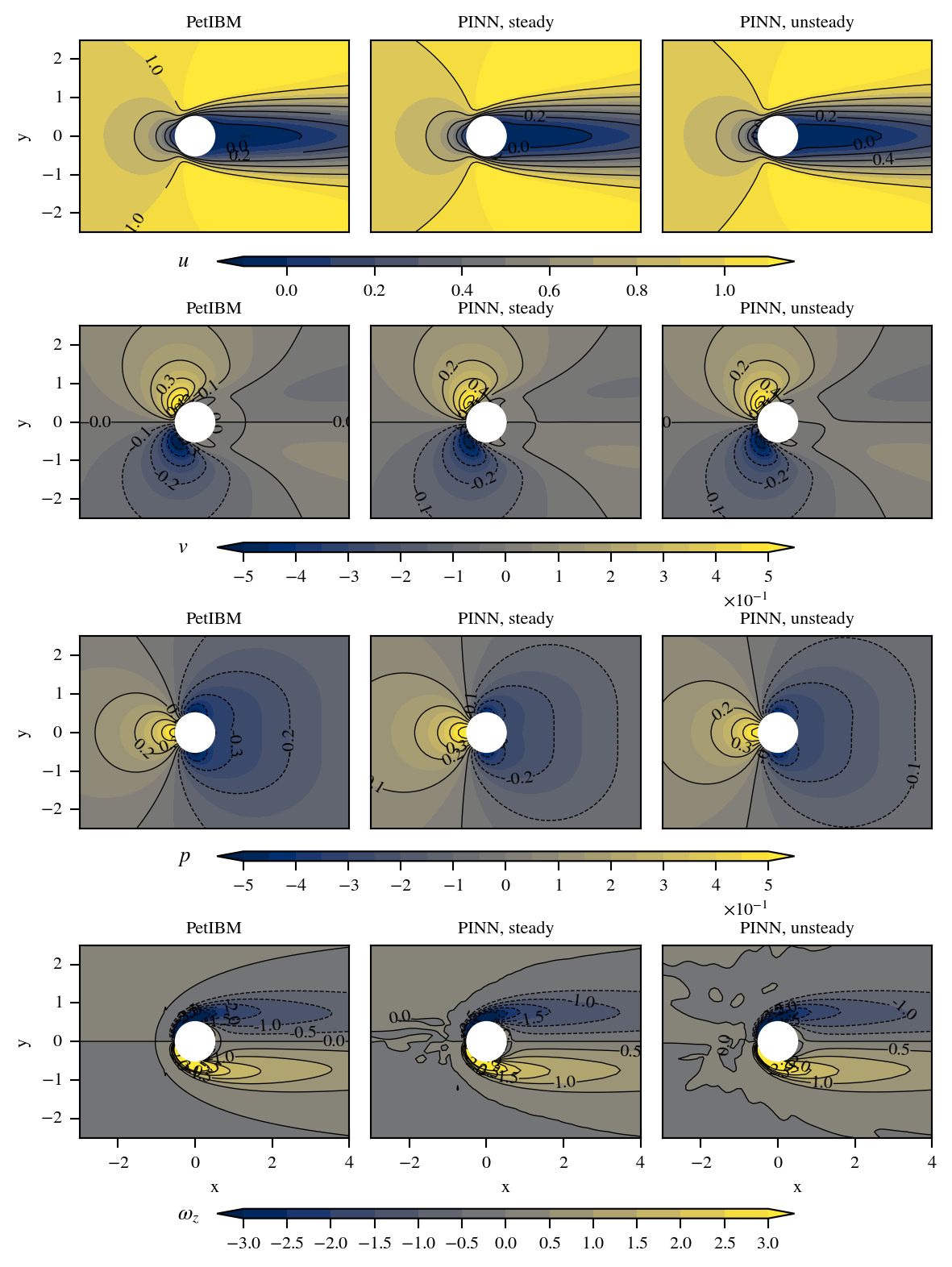}%
    \caption{%
        Contour plots for 2D cylinder flow at $Re=\num{40}$
    }
    \label{fig:cylinder-re40-contours}%
\end{figure*}

Figure \ref{fig:cylinder-re40-pinn-surfp} shows the pressure distribution on the cylinder surface.
Again, though there is not a single solution that all works agree upon, the results from PetIBM and the PINN solvers visually agree with the published literature.
We consider PetIBM and both PINN solvers validated.
Finally, figure \ref{fig:cylinder-re40-contours} compares the steady-state flow fields (i.e., the snapshots at $t=20$ for PetIBM and the unsteady PINN solver).
The PINN solvers' results visually agree with PetIBM's.
The variation in the vorticity of PINNs only happens at the contour line of \num{0}, so it is likely caused by trivial rounding errors.
Note that vorticity is obtained by post-processing for all solvers.
PetIBM used central difference to calculate the vorticity, while the PINN solvers used automatic differentiation to obtain it.

\begin{table}[h]
    \centering%
    \begin{threeparttable}
        \begin{tabular}{lccc}
            \toprule
            & $C_D$ & $C_{D_p}$ & $C_{D_f}$ \\
            \midrule
            Steady PINN & 1.62 & 1.06 & 0.55 \\
            Unsteady PINN & 1.60 & 1.06 & 0.55 \\
            PetIBM & 1.63 & 1.02 & 0.61 \\
            Rosetti et al., 2012\cite{rosetti_urans_2012}\tnote{1} & \num{1.74+-0.09} & n/a & n/a \\
            Rosetti et al., 2012\cite{rosetti_urans_2012}\tnote{2} & 1.61 & n/a & n/a \\
            Sen et al., 2009\cite{sen_steady_2009}\tnote{2} & 1.51 & n/a & n/a \\
            Park et al., 1988\cite{park_numerical_1998}\tnote{2} & 1.51 & 0.99 & 0.53 \\
            Tritton, 1959\cite{tritton_experiments_1959}\tnote{1} & 1.48--1.65 & n/a & n/a \\
            Grove et al., 1964\cite{grove_experimental_1964}\tnote{1} & n/a & 0.94 & n/a \\
            \bottomrule
        \end{tabular}%
        \begin{tablenotes}
            \footnotesize
            \item [1] Experimental result
            \item [2] Simulation result
        \end{tablenotes}
        \caption{%
            Validation of drag coefficients. %
            $C_D$, $C_{D_p}$, and $C_{D_f}$ denote the coefficients of total drag, pressure drag, %
            and friction drag, respectively.%
        }%
        \label{table:cylinder-re40-cd-comparison}
    \end{threeparttable}
\end{table}%


    \section{Case Study: 2D Cylinder Flow at $Re=\num{200}$}\label{sec:case-study}

The previous section presented successful verification with a Taylor-Green vortex having an analytical solution, and validation of the solvers with the $Re=40$ cylinder flow, which exhibits a steady state solution.
Those results give confidence that the solvers are correctly solving the Navier-Stokes equations, and able to model vortical flow. In this section, we study the case of cylinder flow at $Re=200$, exhibiting vortex shedding.

\subsection{Case configurations}

The computational domain is $[-8$, $25]$ $\times$ $[-8$, $8]$ for $x$ and $y$, and $t\in[0$, $200]$.
Other boundary conditions, initial conditions, and density were the same as those in section \ref{sec:val_2d_cylinder_re40}.
The non-dimensional kinematic viscosity is set to $0.005$ to make the Reynolds number $200$.

The PetIBM simulation was done with a grid resolution of $1485$ $\times$ $720$ and $\Delta t = \num{5e-3}$.
The hardware used and the configurations of the linear solvers were the same as described in section \ref{sec:val_2d_cylinder_re40}.

As for the PINN solvers, in addition to the steady and unsteady solvers, a third PINN solver was used: a data-driven PINN.
The data-driven PINN solver is the same as the unsteady PINN solver but replaces the three initial condition losses ($L_4$ to $L_6$) with:

\begin{equation}\label{eq:data-driven-loss}
    \left\{
        \begin{array}{l}
            L_4 = u - u_{data}\\
            L_5 = v - v_{data}\\
            L_6 = p - p_{data}\\
        \end{array}
    \right.
    ,\text{ if }
    \begin{array}{l}
        \vec{x} \in \Omega \\
        t \in T_{data}
    \end{array}
\end{equation}
where subscript $data$ denotes data from a PetIBM simulation.
$T_{data}$ denotes the time range for which we feed the PetIBM simulation data to the data-driven PINN solver.
In this case, $T_{data} \equiv \left[125, 140\right]$.
The PetIBM simulation outputted transient snapshots every 1 second in simulation time, hence the data fed to the data-driven PINN solver consisted of 16 snapshots.
These snapshots contain around $3$ full periods of vortex shedding.
The total number of spatial-temporal points in these snapshots is around $\num{17000000}$, and we only used $\num{6400}$ every iteration, meaning each data batch was repeated approximately every $\num{2650}$ iterations.
Except for replacing the IC losses with a data-driven approach, all other loss terms and the code in the data-driven PINN solver remain the same as the unsteady PINN solver.

Note that for the data-driven PINN solver, the PDE and boundary condition losses were evaluated only in $t\in[125$, $200]$ because we treated the PetIBM snapshots as if they were initial conditions.
Another note is the use of steady PINN solver.
The $Re=200$ cylinder flow is not expected to have a steady-state solution.
However, it is not uncommon to see a steady-state flow solver used for unsteady flows for engineering purposes, especially two or three decades ago when computing power was much more restricted.

The MLP network used on all PINN solvers has 6 hidden layers and 512 neurons per layer.
The configurations of spatial-temporal points are the same as those in section \ref{sec:val_2d_cylinder_re40}.
The Adam optimizer is also the same, except that now we ran for \num{1000000} optimization iterations.
The parameters of the cyclical learning rate scheduler are now: $\eta_{low}=\num{1e-6}$, $\eta_{high}=\num{1e-2}$, $N_c=5000$, and $\gamma=\num{0.9999915}$.
The hardware used was one NVIDIA A100 GPU for all PINN runs.

\subsection{Results}\label{sec:cylinder-re200-results}

The overall run times for the steady, unsteady, and data-driven PINN solvers were about 28 hours, 31 hours, and 33.5 hours using one A100 GPU.
The PetIBM simulation, on the other hand, took around 1.7 hours with a K40 GPU, which is 5-generation-behind in terms of the computing technology.

\begin{figure}
    \centering%
    \includegraphics[width=0.95\columnwidth]{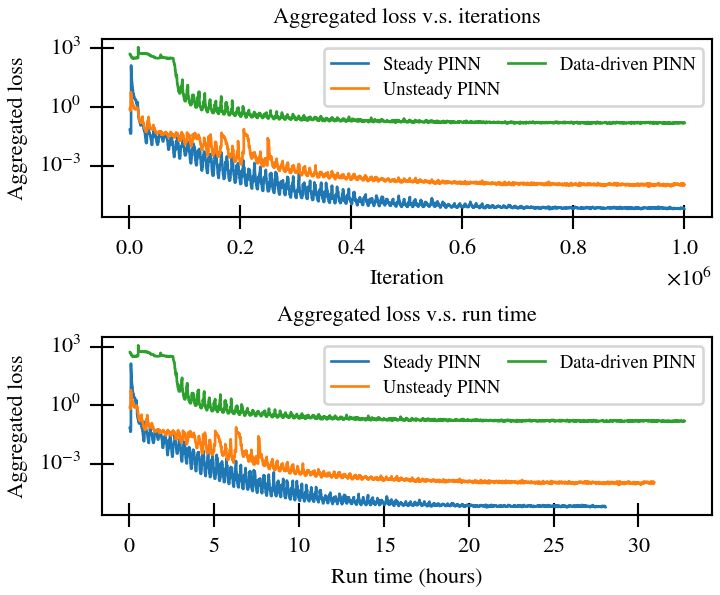}%
    \caption{%
        Training convergence history of 2D cylinder flow at $Re=\num{200}$ w/ PINNs
    }
    \label{fig:cylinder-re200-pinn-loss}%
\end{figure}

Figure \ref{fig:cylinder-re200-pinn-loss} shows the aggregated loss convergence history of all cases.
It shows both the losses and the run times of the three PINN solvers.
As seen in section \ref{sec:val_2d_cylinder_re40}, the unsteady PINN solver converges to a higher total loss than the steady PINN solver does.
Also, the data-driven PINN solver converges to an even higher total loss.
However, it is unclear at this point if having a higher loss means a higher prediction error in data-driven PINN because we replaced the initial condition losses with 16 snapshots from PetIBM and only ran the data-driven PINN solver for $t\in[125, 200]$.

\begin{figure}[t]
    \centering%
    \includegraphics[width=0.95\columnwidth]{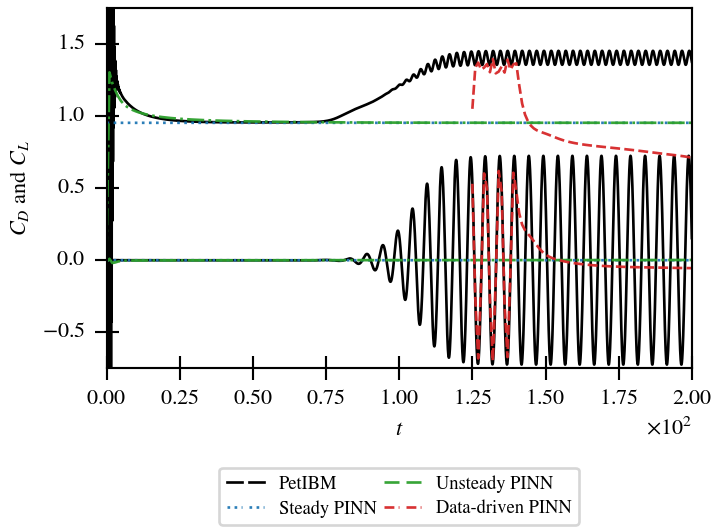}%
    \caption{%
        Drag and lift coefficients of 2D cylinder flow at $Re=\num{200}$ w/ PINNs
    }
    \label{fig:cylinder-re200-drag-lift}%
\end{figure}

Figure \ref{fig:cylinder-re200-drag-lift} shows the drag and lift coefficients versus simulation time.
The coefficients from the steady case are shown as just a horizontal line since there is no time variable in this case.
The unsteady case, to our surprise, does not exhibit oscillations, meaning it results on no vortex shedding, even though it fits well with the PetIBM result before vortex shedding starts (at about $t=75$).
Comparing the coefficients between the steady, unsteady, and PetIBM's values before shedding, we believe the unsteady PINN in this particular case behaves just like a steady solver.
This is supported by the values in table \ref{table:cylinder-2d-re200-cd}, in which we compare $C_D$ against published values in the literature of both unsteady and steady CFD simulations.
\begin{table}
    \centering%
    \begin{threeparttable}[b]
        \begin{tabular}{lcc}
            \toprule
            & $C_D$ \\
            \midrule
            PetIBM & 1.38   \\
            Steady PINN & 0.95 \\
            Unsteady PINN & 0.95 \\
            Deng et al., 2007\cite{deng_hydrodynamic_2007}\tnote{1} & 1.25 \\
            Rajani et al., 2009\cite{Rajani2009}\tnote{1} & 1.34 \\
            Gushchin \& Shchennikov, 1974\cite{gushchin_numerical_1974}\tnote{2} & 0.97 \\
            Fornberg, 1980\cite{fornberg_numerical_1980}\tnote{2} & 0.83 \\
            \bottomrule
        \end{tabular}%
        \begin{tablenotes}
            \footnotesize
            \item [1] Unsteady simulations.
            \item [2] Steady simulations.
        \end{tablenotes}
        \caption{%
            PINNs, 2D Cylinder, $Re=200$: validation of drag coefficients.%
            The data-driven case is excluded because it does not have an obvious periodic state nor a steady-state solution.%
        }%
        \label{table:cylinder-2d-re200-cd}
    \end{threeparttable}
\end{table}%
The $C_D$ obtained from the unsteady PINN is the same as the steady PINN and close to those steady CFD simulations.

As for the data-driven case, its temporal domain is $t\in[125$, $200]$, so the coefficients' trajectories start from $t=125$.
The result, again unexpected to us, only exhibits shedding in the timeframe with PetIBM data, i.e., $t\in[125$, $140]$.
This result also implies that data-driven PINNs may be more difficult to train, compared to data-free PINNs and regular data-only model fitting.
Even in the time range with PetIBM data, the data-driven PINN solver is not able to reach the given maximal $C_L$, and the $C_D$ is obviously off from the given data.
After $t=140$, the trajectories quickly fall back to the no-shedding pattern, though it still deviates from the trajectories of the steady and unsteady PINNs.
Combining the loss magnitude shown in figure \ref{fig:cylinder-re200-pinn-loss}, the deviation of $C_D$ and $C_L$ from the data-driven PINN may be caused by not enough training.
As figure \ref{fig:cylinder-re200-pinn-loss} shows data-driven PINN had already converged, other optimization techniques or hyperparameter tuning may be required to further reduce the loss.
Insufficient training only explains why the data-driven case deviates from the PetIBM's data in $t \in [125, 140]$ and from the other two PINNs for $t > 140$.
Even with a better optimization and eventually a lower loss, based on the trajectories, we do not believe the shedding will continue after $t=140$.

\begin{figure*}
    \centering%
    \includegraphics[width=0.95\textwidth]{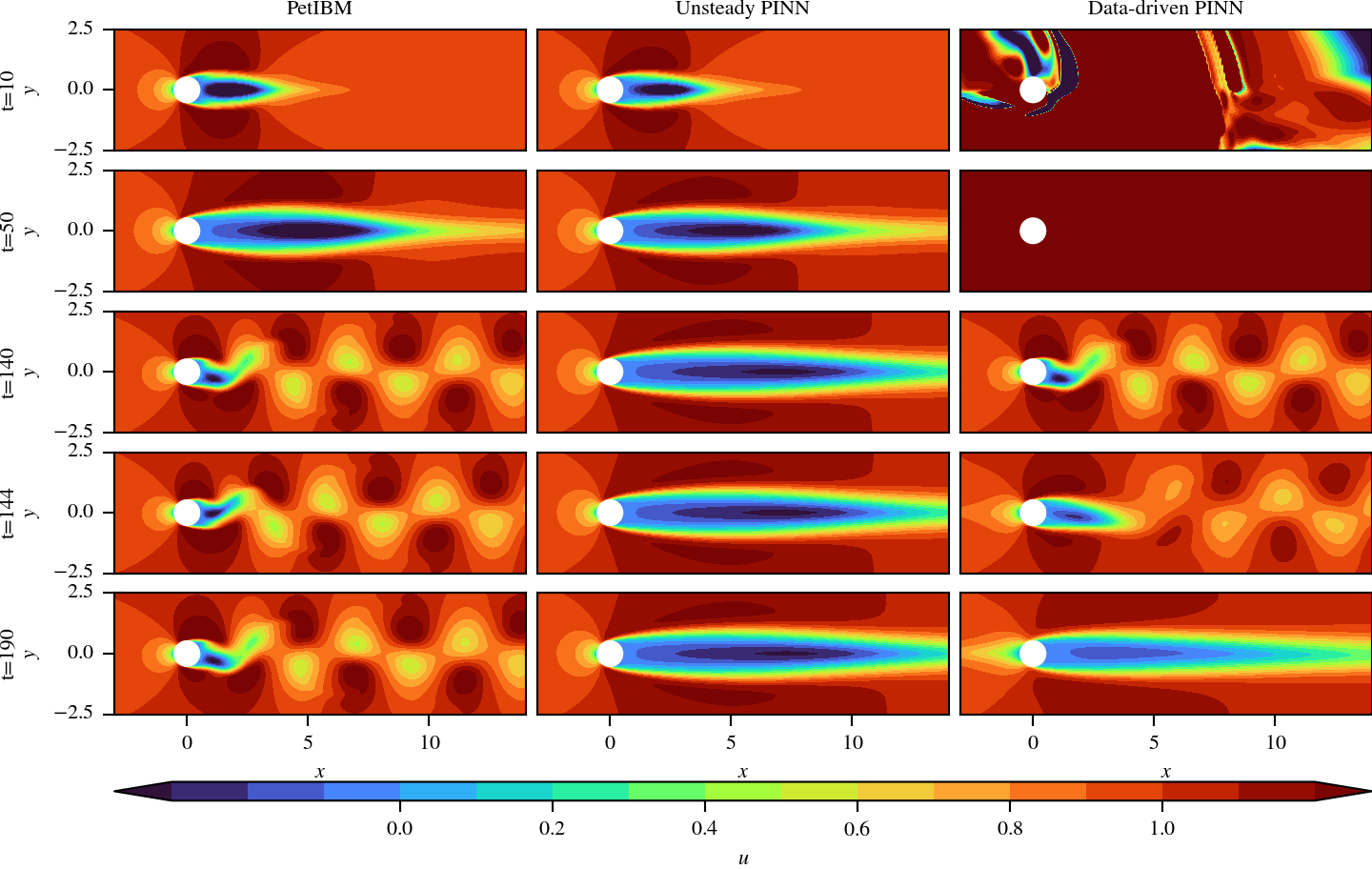}%
    \caption{%
        $u$-velocity comparison of 2D cylinder flow of $Re=\num{200}$ between PetIBM, unsteady PINN, and data-driven PINN.
    }
    \label{fig:cylinder-re200-pinn-contours-u}%
\end{figure*}

\begin{figure*}
    \centering%
    \includegraphics[width=0.95\textwidth]{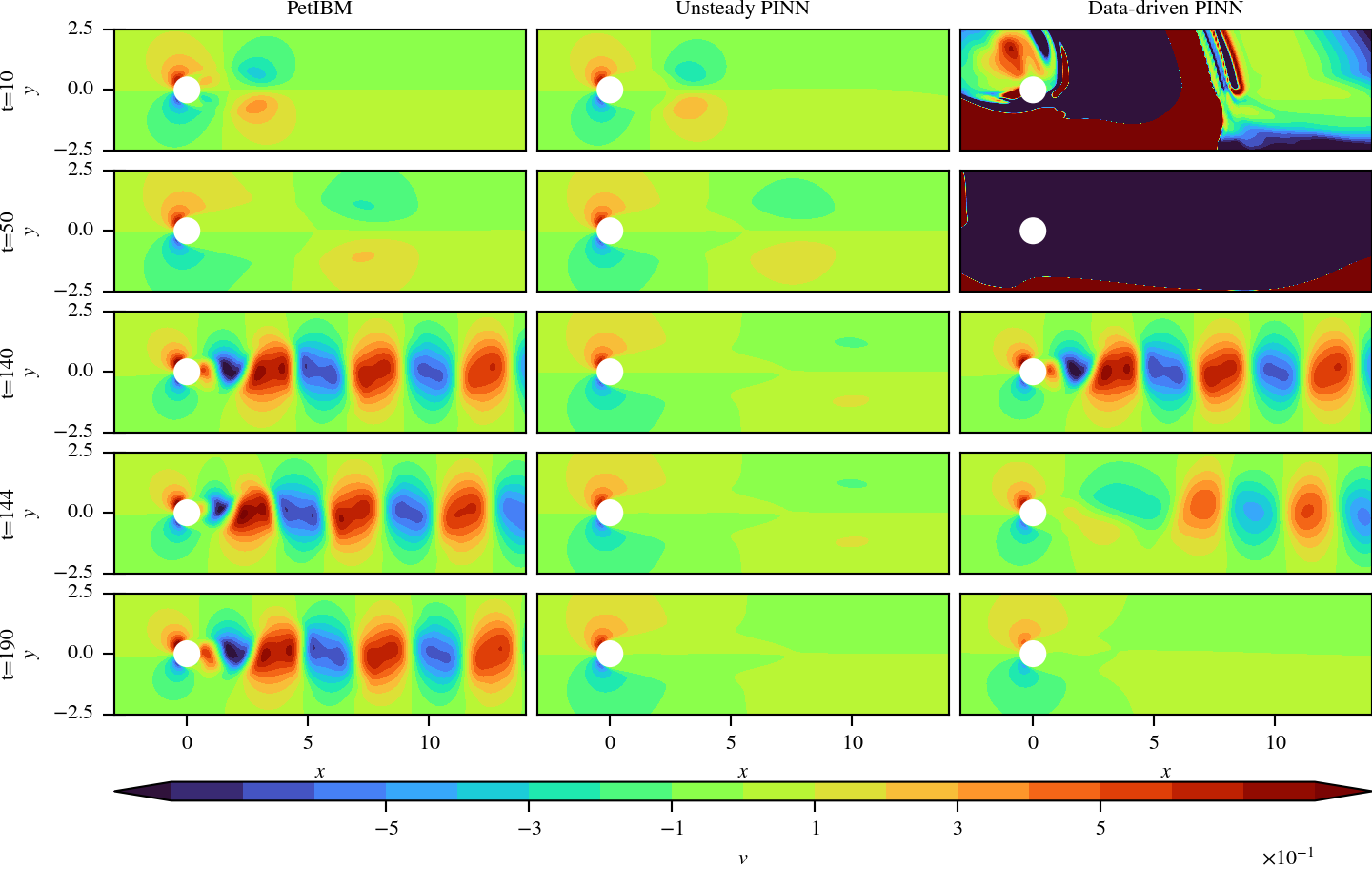}%
    \caption{%
        $v$-velocity comparison of 2D cylinder flow of $Re=\num{200}$ between PetIBM, unsteady PINN, and data-driven PINN.
    }
    \label{fig:cylinder-re200-pinn-contours-v}%
\end{figure*}

\begin{figure*}
    \centering%
    \includegraphics[width=0.95\textwidth]{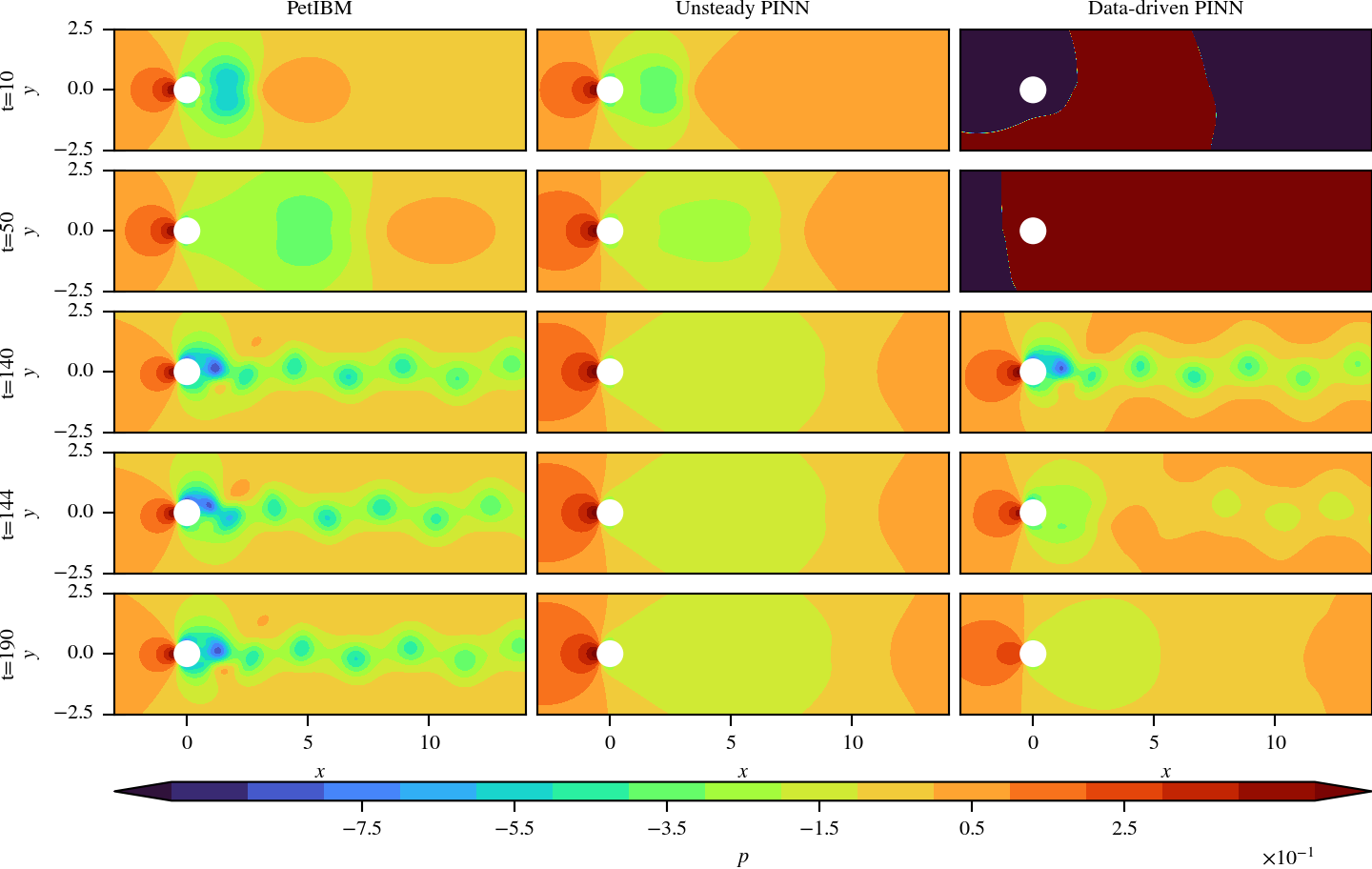}%
    \caption{%
        Pressure comparison of 2D cylinder flow of $Re=\num{200}$ between PetIBM, unsteady PINN, and data-driven PINN.
    }
    \label{fig:cylinder-re200-pinn-contours-p}%
\end{figure*}

\begin{figure*}
    \centering%
    \includegraphics[width=0.95\textwidth]{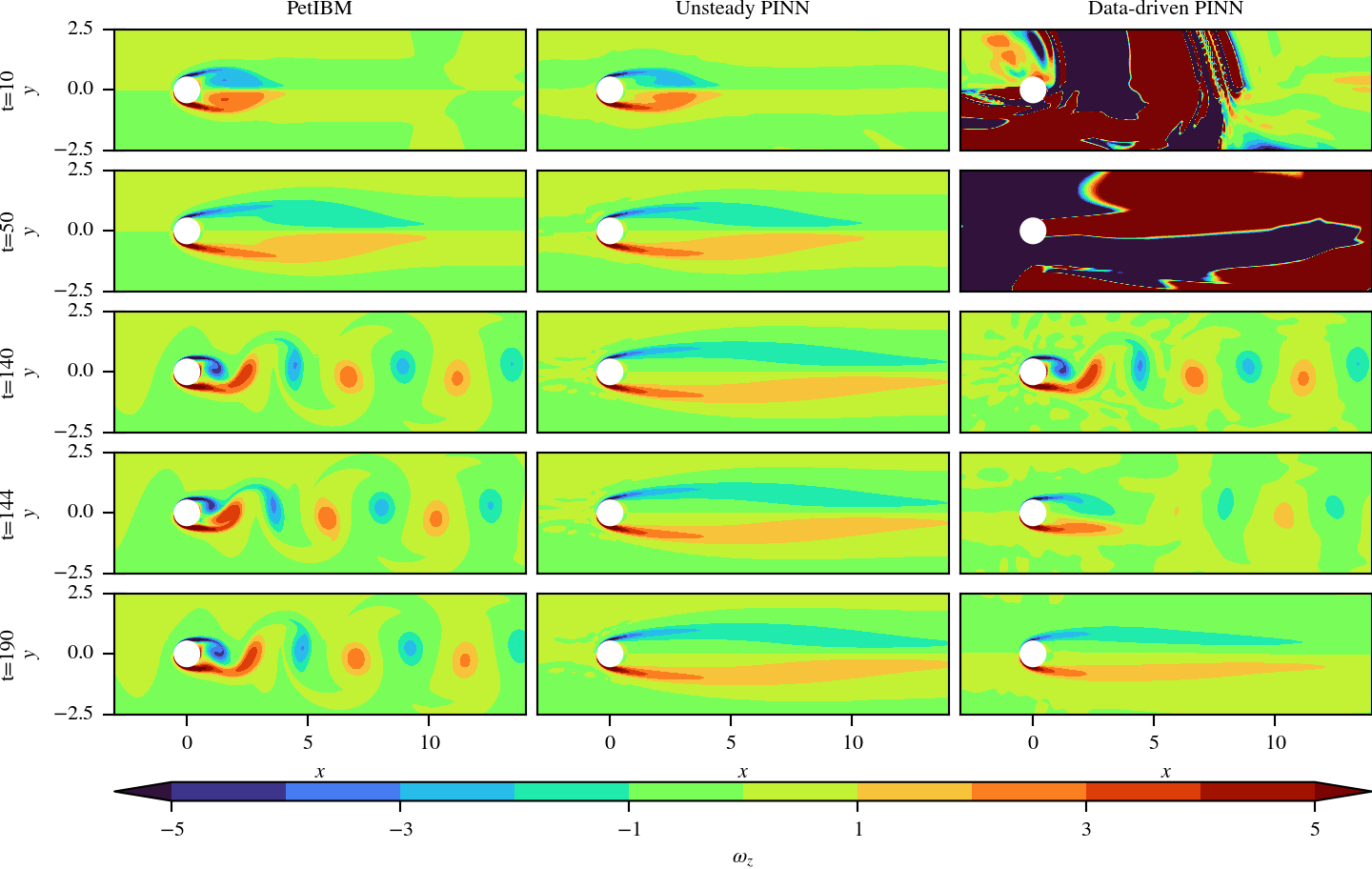}%
    \caption{%
        Vorticity ($\omega_z$) comparison of 2D cylinder flow of $Re=\num{200}$ between PetIBM, unsteady PINN, and data-driven PINN.
    }
    \label{fig:cylinder-re200-pinn-contours-omega_z}%
\end{figure*}

\begin{figure}[t]
    \centering%
    \includegraphics[width=0.95\columnwidth]{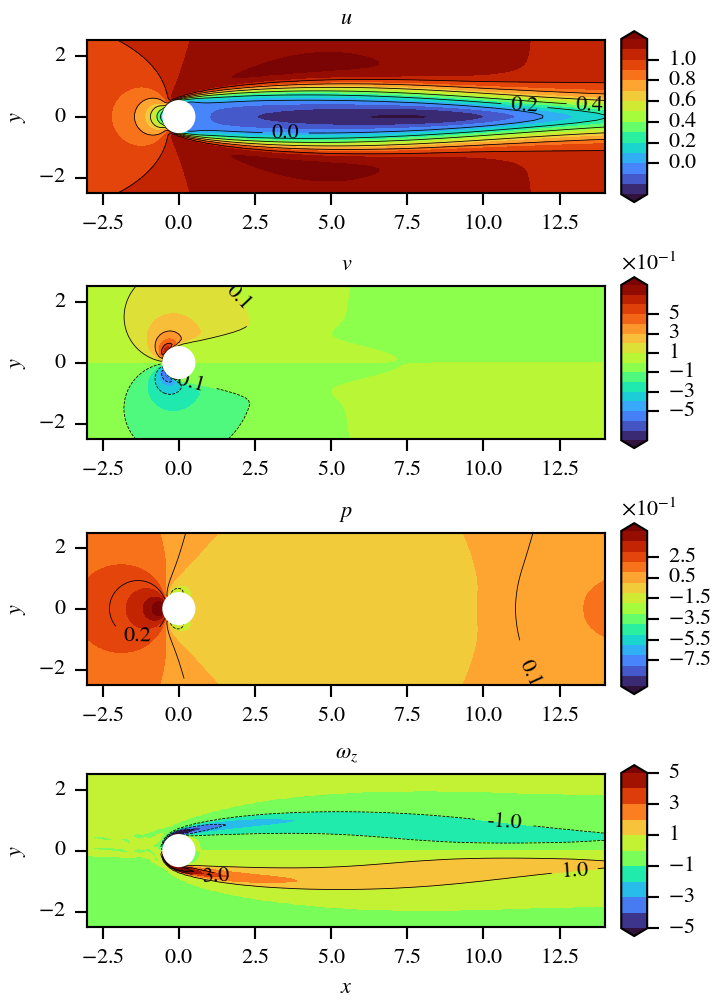}%
    \caption{%
        Contours of 2D cylinder flow at $Re=\num{200}$ w/ steady PINN.
    }
    \label{fig:cylinder-re200-steady-pinn-contours}%
\end{figure}

To examine how the transient flow develops, we visually compared several snapshots of the flow fields from PetIBM, unsteady PINN, and the data-driven PINN, shown in figures \ref{fig:cylinder-re200-pinn-contours-u}, \ref{fig:cylinder-re200-pinn-contours-v}, \ref{fig:cylinder-re200-pinn-contours-p}, and \ref{fig:cylinder-re200-pinn-contours-omega_z}.
We also present the flow contours from the steady PINN in figure \ref{fig:cylinder-re200-steady-pinn-contours} for reference.

At $t=10$, we can see the wake is still developing, and the unsteady PINN visually matches PetIBM.
At $t=50$, the contours again show the unsteady PINN matching the PetIBM simulation before shedding.
These observations verify that the unsteady PINN is indeed solving unsteady governing equations.
The data-driven PINN does not show meaningful results because $t=10$ and $50$ are out of the data-driven PINN's temporal domain.
These results also indicate that the data-driven PINN is not capable of extrapolating backward in time in dynamical systems.

At $t=140$, vortex shedding has already happened.
However, the unsteady PINN solution does not show any shedding.
Moreover, the unsteady PINN's contour plot is similar to the steady case in figure \ref{fig:cylinder-re200-steady-pinn-contours}.
$t=140$ is also the last snapshot we fed into the data-driven PINN for training.
The contour of the data-driven PINN at this time shows that it at least could qualitatively capture the shedding, which is expected.
At $t=144$, it is just $4$ time units from the last snapshot being fed to the data-driven PINN.
At such time, the data-driven PINN has already stopped generating new vortices.
The existing vortex can be seen moving toward the boundary, and the wake is gradually restoring to the steady state wake.
Flow at $t=190$ further confirms that the data-driven PINN's behavior is tending toward that of the unsteady PINN, which behaves like a steady state solver.
On the other hand, the solutions from the unsteady PINN for these times remain steady.

\begin{figure}
    \centering%
    \includegraphics[width=\columnwidth]{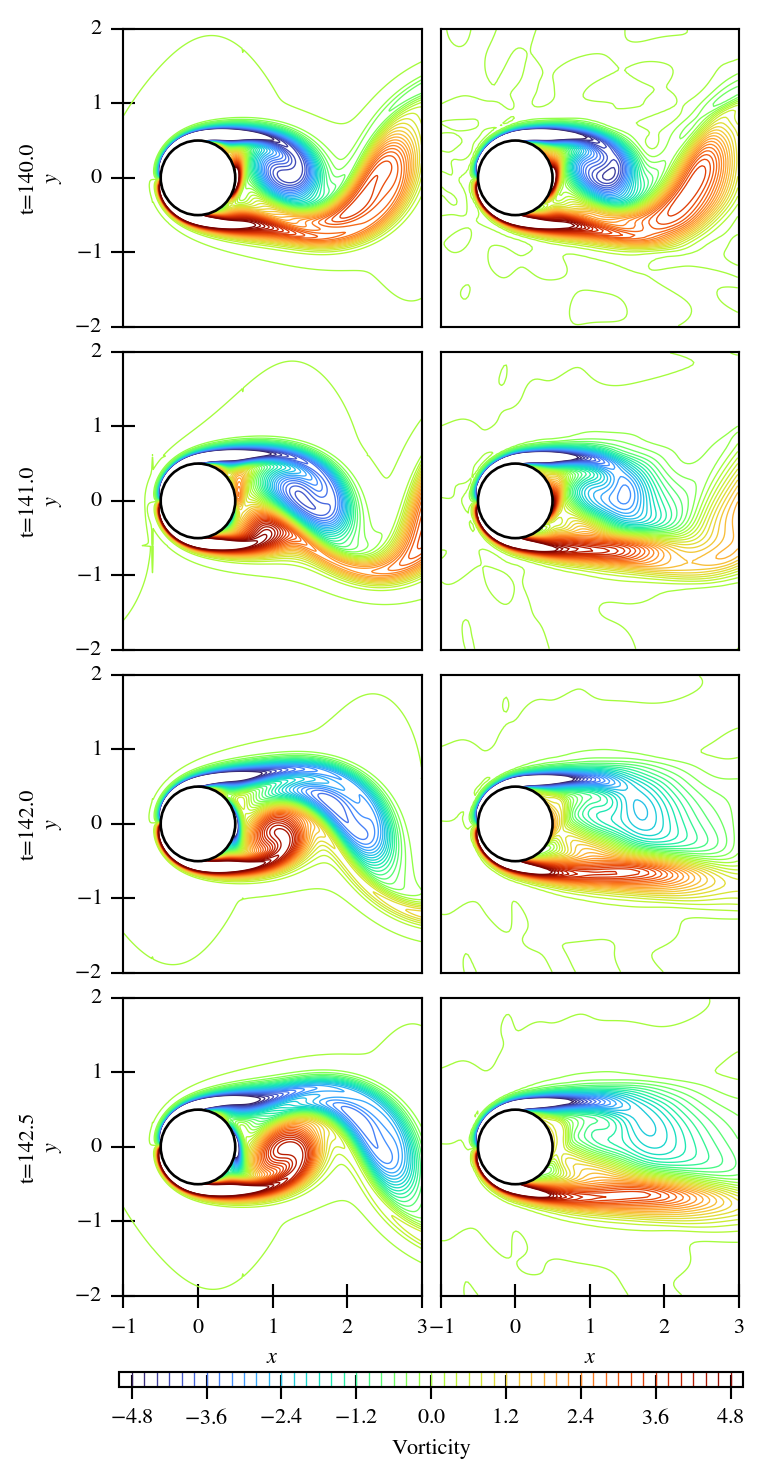}%
    \caption{%
        Vorticity generation near the cylinder for 2D cylinder flow of $Re=\num{200}$ at $t=140$, $141$, $142$, and $142.5$ w/ data-driven PINNs.
    }
    \label{fig:cylinder-re200-pinn-vort-gen}%
\end{figure}

Figure \ref{fig:cylinder-re200-pinn-vort-gen} shows the vorticity from PetIBM and the data-driven PINN in the vicinity of the cylinder in $t \in [140, 142.5]$, which contains a half cycle of vortex shedding.
These contours compare how vorticity was generated right after we stopped feeding PetIBM data into the data-driven PINN.
These comparisons might shed some light on why the data-free PINN cannot generate vortex shedding and why the data-driven PINN stops to do so after $t=140$.

At $t=140$, PetIBM and the data-driven PINN show visually indistinguishable vorticity contours.
This is expected as the data-driven PINN has training data from PetIBM at this time.
At $t=141$ in PetIBM's results, the main clockwise vortex (the blue circular pattern in the domain of $[1, 2]\times[-0.5, 0.5]$) moves downstream.
It slows down the downstream $u$ velocity and accelerates the $v$ velocity in $y<0$.
Intuitively, we can treat the main clockwise vortex as a blocker that blocks the flow in $y<0$ and forces the flow to move upward.
The net effect is the generation of a counterclockwise vortex at around $x\approx 1$ and $y \in [-0.5, 0]$.
This new counterclockwise vortex further generates a small but strong secondary clockwise vortex on the cylinder surface in $y\in[-0.5, 0]$.
On the other hand, the result of the data-driven PINN at $t=141$ shows that the main clockwise vortex becomes more dissipated and weaker, compared to that in PetIBM.
It is possible that the main clockwise vortex is not strong enough to slow down the flow in $y<0$ nor to bring the flow upward.
The downstream flow in $y<0$ (the red arm-like pattern below the cylinder) thus does not change its direction and keeps going straight down in the $x$ direction.
In the results of $t=142$ and $t=142.5$ from PetIBM, the flow completes a half cycle.
That is, the flow pattern at $t=142.5$ is an upside down version of that at $t=140$.
The results from the data-driven PINN, however, do not have any new vortices and the wake becomes more like steady flow.
These observations might indicate that the PINN is either diffusive or dissipative (or both).

\begin{figure}
    \centering%
    \includegraphics[width=\columnwidth]{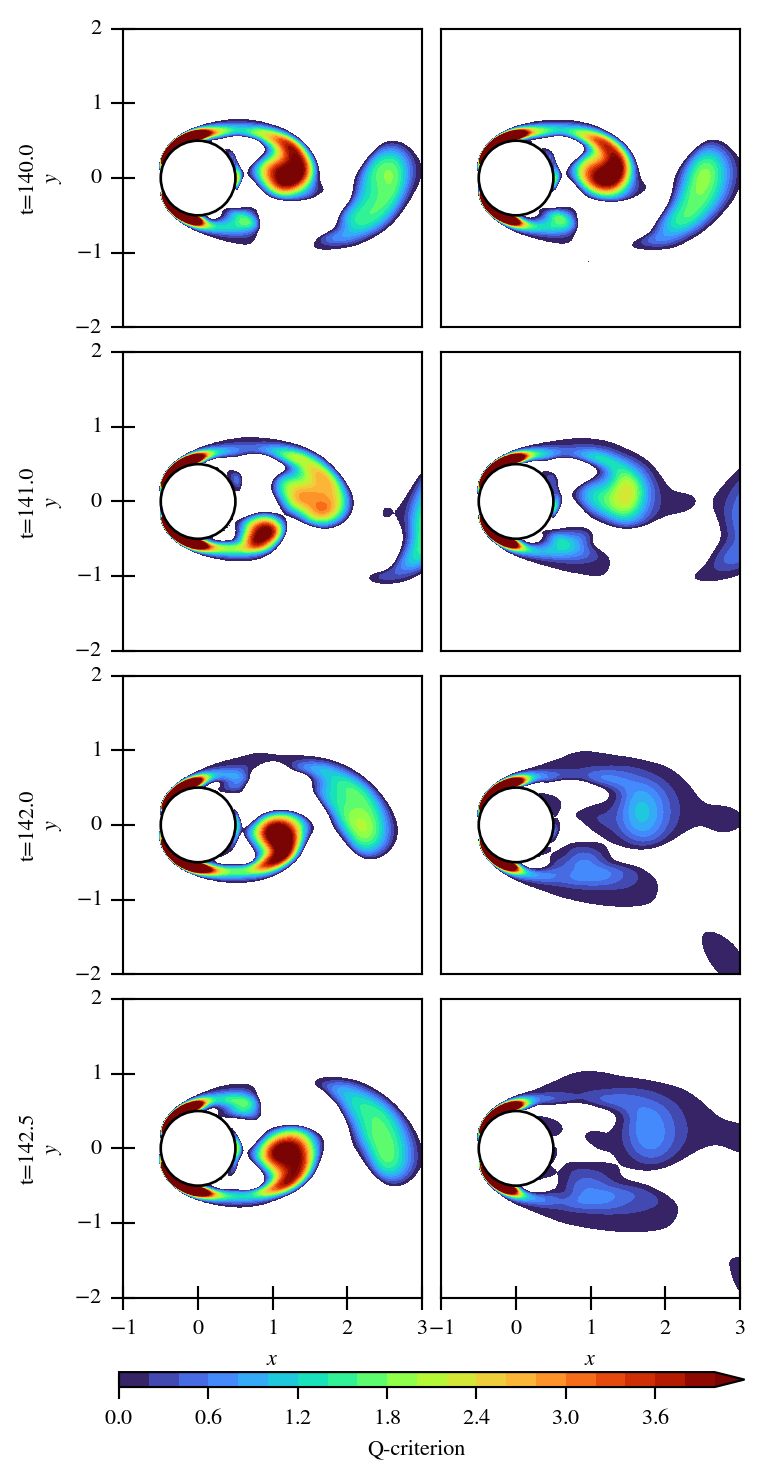}%
    \caption{%
        Q-criterion generation near the cylinder for 2D cylinder flow of $Re=\num{200}$ at $t=140$, $141$, $142$, and $142.5$ w/ data-driven PINNs.
    }
    \label{fig:cylinder-re200-pinn-qcriterion}%
\end{figure}

Next, we examined the Q-criterion in the same vicinity of the cylinder in $t\in[140, 142.5]$, shown in figure \ref{fig:cylinder-re200-pinn-qcriterion}.
The Q-criterion is defined as follows \cite{jeong_identification_1995}:
\begin{equation}
    Q \equiv \frac{1}{2}\left(\lVert \Omega \rVert^2 - \lVert S \rVert^2\right),
\end{equation}
where $\Omega\equiv\frac{1}{2}\left(\nabla\vec{u}-\nabla\vec{u}^\mathsf{T}\right)$ is the vorticity tensor;
$S\equiv\frac{1}{2}\left(\nabla\vec{u}+\nabla\vec{u}^\mathsf{T}\right)$ is the strain rate tensor;
and $\nabla\vec{u}$ is the velocity gradient tensor.
A criterion $Q > 0$ identifies a vortex structure in the fluid flow, that is, where the rotation rate is greater than the strain rate.

We observe that vortices in the data-driven PINN are diffusive and could be dissipative.
Moreover, judging by the locations of vortex centers, vortices also move slower in the PINN solution than with PetIBM.
The edges of the vortices move at a different speed from that of the vortex centers in the PINN case.
This might be hinting at the existence of numerical dispersion in the PINN solver.

\subsection{Dynamical Modes and Koopman Analysis}\label{sec:cylinder-re200-koopman}

We conducted spectral analysis on the cylinder flow to extract frequencies embedded in the simulation results.
Fluid flow is a dynamical system, and how information (or signals) propagates in time plays an important role.
Information with different frequencies advances at different speeds in the temporal direction, and the superposition of information forms complicated flow patterns over time.
Spectral analysis reveals a set of modes, each associated with a fixed oscillation frequencies and decay or growth rate, called {\it dynamic modes} in fluid dynamics.
By comparing the dynamic modes in the solutions obtained with PINNs and PetIBM, we may examine how well or how badly the data-driven PINN learned information with different frequencies.
Koopman analysis is a method to achieve such spectral analysis for dynamical systems.
Please refer to {\it the method of snapshots} in reference \cite{chen_variants_2012} and reference \cite{rowley_spectral_2009} for the theory and the algorithms used in this work.

We analyzed the results from PetIBM and the data-driven PINN in $t\in$$[125$, $140]$, which contains about three full cycles of vortex shedding.
A total of $76$ solution snapshots were used from each solver.
The time spacing is $\Delta t = 0.2$.
The Koopman analysis would result in $75$ modes.
Since the snapshots cover three full cycles, we would expect only $25$ distinct frequencies and $25$ nontrivial modes---only $25$ out of $76$ snapshots are distinct.
However, this expectation only happens when the data are free from noise and numerical errors and when the number of three periods is exact.
We would see more than $25$ distinct frequencies and modes if the data were not ideal.
In $t \in [125, 140]$, the data-driven PINN was trained against PetIBM's data, so we expected to see similar spectral results between the two solvers.

\begin{figure}
    \centering%
    \includegraphics[width=\columnwidth]{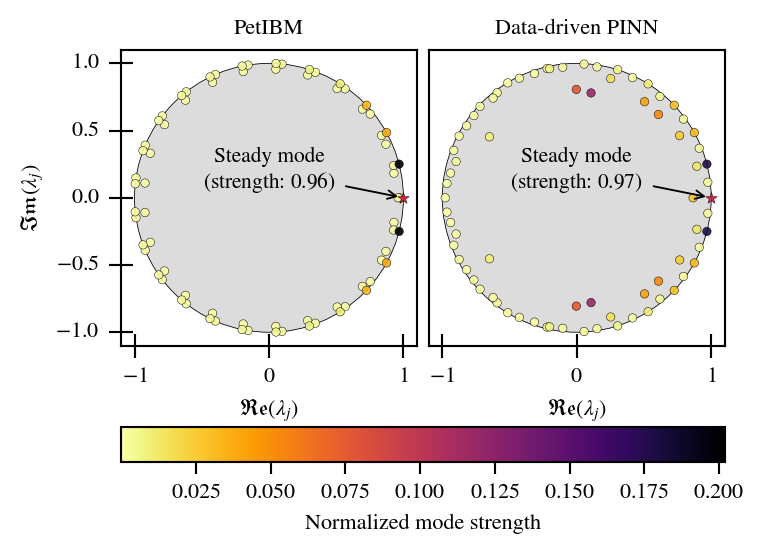}%
    \caption{%
        Distribution of the Koopman eigenvalues on the complex plane for 2D cylinder flow at $Re=\num{200}$ obteined with PetIBM and with data-driven PINN.
    }
    \label{fig:cylinder-re200-koopman-eig-dist}%
\end{figure}

To put it simply, each dynamic mode is identified by a complex number.
Taking logarithm on the complex number's absolute value gives a mode's growth rate, and the angle of the complex number corresponds to a mode's frequency.
Figure \ref{fig:cylinder-re200-koopman-eig-dist} shows the distributions of the dynamic modes on the complex plane.
The color of each dot represents the normalized strength of the corresponding mode, which is also obtained from the Koopman analysis.
The star marker denotes the mode with a frequency of zero, i.e., a steady or time-independent mode.
This mode usually has much higher strength than others, so we excluded it from the color map and annotated its strength directly.
Koopman analysis delivers dynamical modes with complex conjugate pairs, so the modes are symmetric with respect to the real-number axis. 
A conjugate pair has an opposite sign in the frequencies mathematically but has the same physical frequency.

We also plotted a circle with a radius of one on each figure.
As flow has already reached a fully periodic regime, the growth rates should be zero because no mode becomes stronger nor weaker.
In other words, all modes were expected to fall on this circle on the complex plane.
If a mode falls inside the circle, it has a negative growth rate, and its contribution to the solution diminishes over time.
Similarly, a mode falling outside the circle has a positive growth rate and becomes stronger over time.

On the complex plane, all the modes captured by PetIBM (the left plot in figure \ref{fig:cylinder-re200-koopman-eig-dist}) fall onto the circle or very close to the circle.
The plot shows $75$---rather than $25$---distinct $\lambda_j$ and modes, but the modes are evenly clustered into 25 groups.
Each group has three modes, among which one or two modes fall on the circle, while the remaining one(s) falls inside but very close to the circle.
Modes within each group have a similar frequency, and the one precisely on the circle has significantly higher strength than other modes (if not all modes in the group are weak).
Due to the numerical errors in PetIBM's solutions, data in a vortex period are similar to but not exactly the same as those in another period.
The strong modes falling precisely on the circle may represent the period-averaged flow patterns and are the $25$ modes we expected earlier. 
The effect of numerical errors was filtered out from these modes.
We call these 25 modes primary modes and all other modes secondary modes.
Secondary modes are mostly weak and may come from the numerical errors in the PetIBM simulation.
The plot shows these secondary modes are slightly dispersive but non-increasing over time, which is reasonable because the numerical schemes in PetIBM are stable.

As for the PINN result (the right pane in figure \ref{fig:cylinder-re200-koopman-eig-dist}), the mode distribution is not as structured as with PetIBM.
It is hard to distinguish if all 25 expected modes also exist in this plot.
However, we observe that at least the top 7 primary modes (the steady mode, two purple and 4 orange dots on the circle) also exist in the PINN case.
Secondary modes spread out more widely on and inside the circle, compared to the clustered modes in PetIBM.
We believe this means that PINN is more numerically dispersive and noisy.
The frequencies of many of these secondary modes do not exist in PetIBM.
So one possible source of these additional frequencies and modes may be the PINN method itself.
It could be insufficient training or that the neural network itself inherently is dispersive. 
However, secondary modes on the circle are weak.
We suspect that their contribution to the solution may be trivial.

A more concerning observation is the presence of damped modes (modes that fall inside the circle). 
These modes have negative growth rates and hence are damped over time.
We believe these modes contribute significantly to the solution because their strengths are substantial.
The existence of the damped modes also means that PINN's predictions have more important discrepancies from one vortex period to another vortex period, compared to the PetIBM simulation.
In addition, the flow pattern in PINN would keep changing after $t=140$.
They may be the culprits causing the PINN solution to quickly fall back to a non-oscillating flow pattern for $t>140$.
We may consider these errors as numerical dissipation.
However, whether these errors came from insufficient training or were inherent in the PINN is unclear.

Note that the spectral analysis was done against data in $t\in[125, 140]$.
It does not mean the solutions in $t>140$ also have the same spectral characteristics: the flow system is nonlinear, but the Koopman analysis uses linear approximations \cite{rowley_spectral_2009}.

\begin{figure}
    \centering%
    \includegraphics[width=\columnwidth]{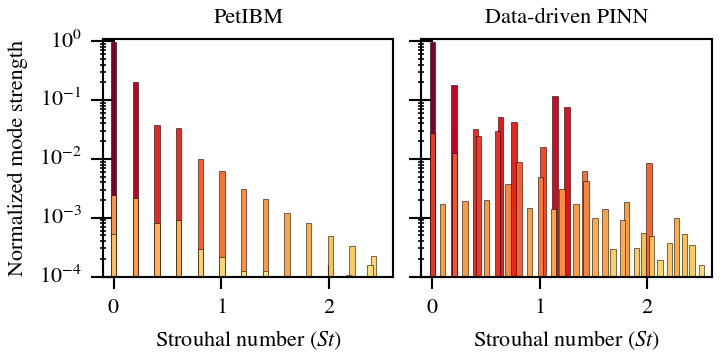}%
    \caption{%
        Mode strengths versus mode frequencies for 2D cylinder flow at $Re=\num{200}$.
        Note that we use a log scale for the vertical axis.
    }
    \label{fig:cylinder-re200-koopman-mode-strength}%
\end{figure}

Figure \ref{fig:cylinder-re200-koopman-mode-strength} shows mode strengths versus frequencies.
The plots use nondimensional frequency, i.e., Strouhal number, in the horizontal axes. 
We only plotted modes with positive numerical frequencies for a concise visualization.
Plots in this figure also show the same observations as in the previous paragraphs: the data-driven PINN is more dispersive and dissipative.

An observation that is now clearer from figure \ref{fig:cylinder-re200-koopman-mode-strength} is the strength distribution.
In PetIBM's case, strengths decrease exponentially from the steady mode (i.e., $St=0$) to high-frequency modes.
One can deduce a similar conclusion from PetIBM's simulation result.
The vortex shedding is dominated by a single frequency (this frequency is $St\approx 0.2$ because $t\in[125, 140]$ contains three periods).
Therefore, the flow should be dominated by the steady mode and a mode with a frequency close to $St=0.2$.
We can indeed verify this statement for PetIBM's case in figure \ref{fig:cylinder-re200-koopman-mode-strength}: the primary modes of $St=0$ and $St\approx 0.2$ are much stronger than others.
The strength of the immediately next mode, i.e., $St\approx 0.4$, drops by an order of magnitude. 
Note the use of a logarithmic scale.
If we re-plot the figure using a regular scale, only $St=0$ and $St=0.2$ would be visible in the figure.

The strength distribution in the case of PINN also shows that $St=0$ and $St\approx 0.2$ are strong.
However, they are not the only dominating modes.  
Some other modes also have strengths at around \num{e-1}.
As discussed in the previous paragraphs, these additional strong modes are damped modes.
We also observed that some damped modes have the same frequencies as primary modes.
For example, the secondary modes at $St=0$ and $St=0.2$ are damped modes.
Note that for $St=0$, if a mode is damped, then it is not a steady mode anymore because its magnitude changes with time, though it is still non-oscillating.

Table \ref{table:koopman-petibm} summarizes the top 4 modes (ranked by their strengths) in PetIBM's spectral result.
\begin{table}
    \begin{threeparttable}[b]
        \begin{tabular}{ccccc}
            \toprule
            $St$ & Strength & Growth Rate & Contours \\
            \midrule
            0     & 0.96 & 1.3e-7  & Figure \ref{fig:cylinder-re200-koopman-petibm-1st}\\
            0.201 & 0.20 & -4.3e-7 & Figure \ref{fig:cylinder-re200-koopman-petibm-2nd}\\
            0.403 & 0.04 & 1.7e-6  & Figure \ref{fig:cylinder-re200-koopman-petibm-3rd}\\
            0.604 & 0.03 & 2.7e-6  & Figure \ref{fig:cylinder-re200-koopman-petibm-4th}\\
            \bottomrule
        \end{tabular}%
        \caption{%
            2D Cylinder, $Re=200$: top 4 primary dynamic modes (sorted by strengths) for PetIBM%
        }%
        \label{table:koopman-petibm}
    \end{threeparttable}
\end{table}%
For reference, these modes' contours are provided in the appendex as denoted in the table.
The dynamic modes are complex-valued, and the contours include both the real and the imaginary parts.
Note the growth rates of these 4 modes are not exactly zero but around \num{e-6} and \num{e-7}.
We were unsure if we could treat them as zero at these orders of magnitude.
If not, and if they do cause the primary modes to be slightly damped or augmented over time, then we believe they also serve as a reasonable explanation for the existence of the other 50 non-primary modes in PetIBM---to compensate for the loss or the gain in the primary modes.

Table \ref{table:koopman-pinn-primary} lists the PINN solution's top 4 primary modes, which are the same as those in table \ref{table:koopman-petibm}.
\begin{table}
    \begin{threeparttable}[b]
        \begin{tabular}{ccccc}
            \toprule
            $St$ & Strength & Growth Rate & Contours \\
            \midrule
            0     & 0.97 & -2.2e-6  & Figure \ref{fig:cylinder-re200-koopman-pinn-primary-1st}\\
            0.201 & 0.18 & -9.4e-6  & Figure \ref{fig:cylinder-re200-koopman-pinn-primary-2nd}\\
            0.403 & 0.03 &  2.3e-5  & Figure \ref{fig:cylinder-re200-koopman-pinn-primary-3rd}\\
            0.604 & 0.03 & -8.6e-5  & Figure \ref{fig:cylinder-re200-koopman-pinn-primary-4th}\\
            \bottomrule
        \end{tabular}%
        \caption{%
            2D Cylinder, $Re=200$: top 4 primary dynamic modes (sorted by strengths) for PINN%
        }%
        \label{table:koopman-pinn-primary}
    \end{threeparttable}
\end{table}%
Table \ref{table:koopman-pinn-damped} shows the top 4 secondary modes in the PINN method's result.
\begin{table}
    \begin{threeparttable}[b]
        \begin{tabular}{ccccc}
            \toprule
            $St$ & Strength & Growth Rate & Contours \\
            \midrule
            1.142 & 0.12 & -0.24 & Figure \ref{fig:cylinder-re200-koopman-pinn-damped-1st}\\
            1.253 & 0.08 & -0.22 & Figure \ref{fig:cylinder-re200-koopman-pinn-damped-2nd}\\
            0.633 & 0.05 & -0.14 & Figure \ref{fig:cylinder-re200-koopman-pinn-damped-3rd}\\
            0.761 & 0.04 & -0.13 & Figure \ref{fig:cylinder-re200-koopman-pinn-damped-4th}\\
            \bottomrule
        \end{tabular}%
        \caption{%
            2D Cylinder, $Re=200$: top 4 damped dynamic modes (sorted by strengths) for PINN%
        }%
        \label{table:koopman-pinn-damped}
    \end{threeparttable}
\end{table}%
Corresponding contours are also included in the appendix and denoted in the tables for readers' reference.
The growth rates of the primary modes in the PINN method's result are around \num{e-5} and \num{e-6}, slightly larger than those of PetIBM.
If these orders of magnitude can not be deemed as zero, then these primary modes are slightly damped and dissipative, though the major source of the numerical dissipation may still be the secondary modes in table \ref{table:koopman-pinn-damped}.


    \section{Discussion}

This case study raises significant concerns about the ability of the PINN method to predict flows with instabilities, specifically vortex shedding.
In the real world, vortex shedding is triggered by natural perturbations.
In traditional numerical simulations, however, the shedding is triggered by various numerical noises, including rounding and truncation errors.
These numerical noises mimic natural perturbations.
Therefore, a steady solution could be physically valid for cylinder flow at $Re = 200$ in a perfect world with no numerical noise.
As PINNs are also subject to numerical noise, we expected to observe vortex shedding in the simulations, but the results show that instead the data-free unsteady PINN converged to a steady-state solution.
Even the data-driven PINN reverted back to a steady-state solution beyond the timeframe that was fed with PetIBM's data.
It is unlikely that the steady-state behavior has to do with perturbations.
In traditional numerical simulations, it is sometimes challenging to induce vortex shedding, particularly in symmetrical computational domains.
However, we can still trigger shedding by incorporating non-uniform initial conditions, which serve as perturbations to the steady state solution.
In the data-driven PINN, the training data from PetIBM can be considered as such non-uniform initial conditions.
The vortex shedding already exists in the training data, yet it did not continue beyond the period of data input, indicating that the perturbation is not the primary factor responsible for the steady-state behavior.
This suggests that PINNs have a different reason for their inability to generate vortex shedding compared to traditional CFD solvers.
Other results in the literature that show the two-dimensional cylinder wake \cite{jin_nsfnets_2021} in fact are using high-fidelity DNS data to provide boundary and initial data for the PINN model.
The failure to capture vortex shedding in the data-free mode of PINN was confirmed in recent work by Rohrhofer et al. \cite{rohrhofer_fixedpoints_2023}.

The steady-state behavior of the PINN solutions may be attributed to spectral bias.
Rahaman et al. \cite{rahaman_spectral_2019} showed that neural networks exhibit spectral bias, meaning they tend to prioritize learning low-frequency patterns in the training data.
For cylinder flow, the lowest frequency corresponds to Strouhal number $St=0$.
The data-free unsteady PINN may be prioritizing learning the mode at $St=0$ (i.e., the steady mode) from the Navier-Stokes equations.
The same may apply to the data-driven PINN beyond the timeframe with training data from PetIBM, resulting in a rapid restoration to the non-oscillating solution.
Even within the timeframe with the PetIBM training data, the data-driven PINN may prioritize learning the $St=0$ mode in PetIBM's data.
Although the vortex shedding in PetIBM's data forces the PINN to learn higher-frequency modes to some extent, the shedding modes are generally more difficult to learn due to the spectral bias.
This claim is supported by the history of the drag and lift coefficients of the data-driven PINN (the red dashed line in figure \ref{fig:cylinder-re200-drag-lift}), which was still unable to predict the peak values in $t \in \left[125, 140\right]$, despite extensive training.

The suspicion of spectral bias prompted us to conduct spectral analysis by obtaining Koopman modes, presented in section \ref{sec:cylinder-re200-koopman}.
The Koopman analysis results are consistent with the existence of spectral bias: the data-driven PINN is not able to learn discrete frequencies well, even when trained with PetIBM's data that contain modes with discrete frequencies.

The Koopman analysis on the data-driven PINN's prediction reveals many additional frequencies that do not exist in the training data from PetIBM, and many damped modes that have a damping effect and reduce or prohibit oscillation.
These damped modes may be the cause of the solution restoring to a steady-state flow beyond the timeframe with PetIBM's data.

From a numerical-method perspective, the Koopman analysis shows that the PINN methods in our work are dissipative and dispersive.
The Q-criterion result (figure \ref{fig:cylinder-re200-pinn-qcriterion}) also demonstrates dissipative behavior, which inhibits oscillation and instabilities.
Dispersion can also contribute to the reduction of oscillation strength.
However, it is unclear whether dispersion and dissipation are intrinsic numerical properties or whether we did not train the PINNs sufficiently, even though the aggregated loss had converged (figure \ref{fig:cylinder-re200-pinn-loss}).
Unfortunately, limited computing resources prevented us from continuing the training---already taking orders of magnitude longer than the traditional CFD solver.
More theoretical work may be necessary to study the intrinsic numerical properties of PINNs beyond computational experiments.

Another point worth discussing is the generalizability of data-driven PINNs.
Our case study demonstrates that data-driven PINNs may not perform well when predicting data they have not seen during training, as illustrated by the unphysical predictions generated for $t = 10$ and $t = 50$ in figures \ref{fig:cylinder-re200-pinn-contours-u}, \ref{fig:cylinder-re200-pinn-contours-v}, \ref{fig:cylinder-re200-pinn-contours-p}, and \ref{fig:cylinder-re200-pinn-contours-omega_z}.
While data-driven PINNs are believed to have the advantage of performing extrapolation in a meaningful way by leveraging existing data and physical laws, our results suggest that this ``extrapolation'' capability may be limited.
In data-driven approaches, the training data typically consists of observation data (e.g., experimental or simulation data) and pure spatial-temporal points.
The ``extrapolation'' capability is therefore constrained to the coordinates seen during training, rather than arbitrary coordinates beyond the observation data.

For example, in our case study, $t \in [0, 125]$ corresponds to spatial-temporal points that were never seen during training, $t \in [125, 140]$ contains observation data, and $t \in [140, 200]$ corresponds to spatial-temporal points seen during training but without observation data.
The PINN method's prediction for $t \in [125, 140]$ is considered interpolation.
Even if we accept the steady-state solution as physically valid, then the data-driven PINN can only extrapolate for $t \in [140, 200]$, and fails to extrapolate for $t \in [0, 125]$.
This limitation means that the PINN method can only extrapolate on coordinates it has seen during training.
If the steady-state solution is deemed unacceptable, then the data-driven PINN lacks extrapolation capability altogether and is limited to interpolation.
This raises the interesting research question of how data-driven PINNs compare to traditional deep learning approaches (i.e., those not using PDEs for losses), particularly in terms of performance and accuracy benefits.

It is worth noting that Cai et al. \cite{cai_physics-informed_2021} argue that data-driven PINNs are useful in scenarios where only sparse observation data are available, such as when an experiment only measures flow properties at a few locations, or when a simulation only saves transient data at a coarse-grid level in space and time.
In such cases, data-driven PINNs may outperform traditional deep learning approaches, which typically require more data for training.
However, as we discussed in our previous work \cite{chuang_experience_2022}, using PDEs as loss functions is computationally expensive, increasing the overall computational graph exponentially.
Thus, even in the context of interpolation problems under sparse observation data, the research question of how much additional accuracy can be gained at what cost in computational expense remains an open and interesting question.

Other works have brought up concerns about the limitations of PINN methods in certain scenarios, like flows with shocks \cite{fuks_limitations_2020} and flows with fast variations \cite{krishnapriyan_failure_2021}.
These researchers suggested that the optimization process on the complex landscape of the loss function may be the cause of the failure.
And other works have also highlighted the performance penalty of PINNs compared to traditional numerical methods \cite{grossmann_pinnvsfem_2023}.
In comparison with finite element methods, PINNs were found to be orders of magnitude slower in terms of wall-clock time.
We also observed a similar performance penalty in our case study, where the PINN method took orders of magnitude longer to train than the traditional CFD solver.
We purposely used a very old GPU (NVIDIA Tesla K40) with PetIBM, running on our lab-assembled workstation, while the PINN method was run on a modern GPU (NVIDIA Tesla A100) on a high-performance computing cluster.
However, we did not conduct a thorough performance comparison.
It is unclear what a ``fair'' performance comparison would look like, as the factors affecting runtime are so different between the two methods.

An interesting third option was proposed recently, where the discretized form of the differential equations is used in the loss function, rather than the differential equations themselves \cite{karnakov_odil_2022}.
This approach foregoes the neural-network representation altogether, as the unknowns are the solution values themselves on a discretization grid. 
It shares with PINNs the features of solving a gradient-based optimization problem, taking advantae of automatic differentiation, and being easily implemented in a few lines of code thanks to modern ML frameworks. 
But it does not suffer from the performance penalty of PINNs, showing an advantage of several orders of magnitude in terms of wall-clock time.
Given that this approach uses a completely different loss function, it supports the claims of other researchers that the loss-function landscape is the source of problems for PINNs.


    \section{Conclusion}

In this study, we aimed to expand upon our previous work \cite{chuang_experience_2022} by exploring the effectiveness of physics-informed neural networks (PINNs) in predicting vortex shedding in a 2D cylinder flow at $Re = 200$.
It should be noted that our focus is limited to forward problems involving non-parameterized, incompressible Navier-Stokes equations.

To ensure the correctness of our results, we verified and validated all involved solvers.
Aside from using as a baseline results obtained with PetIBM, we used three PINN solvers in the case study: a steady data-free PINN, an unsteady data-free PINN, and a data-driven PINN.
Our results indicate that while both data-free PINNs produced steady-state solutions similar to traditional CFD solvers, they failed to predict vortex shedding in unsteady flow situations.
On the other hand, the data-driven PINN predicted vortex shedding only within the timeframe where PetIBM training data were available, and beyond this timeframe the prediction quickly reverted to the steady-state solution.
Additionally, the data-driven PINN showed limited extrapolation capabilities and produced meaningless predictions at unseen coordinates.
Our Koopman analysis suggests that PINN methods may be dissipative and dispersive, which inhibits oscillation and causes the computed flow to return to a steady state.
This analysis is also consistent with the observation of a spectral bias inherent in neural networks \cite{rahaman_spectral_2019}.

One interesting research question that arises from our findings is how the cost-performance ratio of data-driven PINNs compares to classical deep learning approaches.
While data-free PINNs are commonly considered as numerical methods for solving PDEs, data-driven PINNs are more akin to supervised machine/deep learning.
However, data-free PINNs have been shown to have inferior cost-performance ratios compared to traditional numerical methods for PDEs (in terms of forward and non-parameterized problems).
The literature suggests that PINNs are best utilized in a data-driven configuration, rather than data-free settings.
Therefore, it would be valuable to quantitatively compare the benefits of data-driven PINNs to those of classical deep learning approaches and understand the associated cost-performance trade-offs.


    \section{Reproducibility statement}

In our work, we strive for achieving reproducibility of the results, and all the code we developed for this research is available publicly on GitHub under an open-source license, while all the data is available in open archival repositories.
PetIBM is an open-source CFD library based on the immersed boundary method, and is available at \url{https://github.com/barbagroup/PetIBM} under the permissive BSD-3 license. 
The software was peer reviewed and published in the Journal of Open Source Software \cite{chuang_petibm_2018}. 
Our PINN solvers based on the NVIDIA \emph{Modulus} toolkit can be found following the links in the GitHub repository for this paper, located at \url{https://github.com/barbagroup/jcs_paper_pinn/}. 
There, the folder prefixed by \texttt{repro-pack} corresponds to a git submodule pointing to the relevant commit on a branch of the repository for the full reproducibility package of the first author's PhD dissertation \cite{chuang_thesis_2023}.
The branch named \texttt{jcs-paper} contains the modified plotting scripts to produce the publication-quality figures in this paper.    
A snapshot of the repro-pack is archived on Zenodo, and the DOI is 10.5281/zenodo.7988067.
As described in the README of the repro-pack, readers can use pre-generated data for plotting the figures in this paper, or they can re-run the solutions using the code and data available in the repro-pack.
The latter option is of course limited by the computational resources available to the reader.
For the first option, the reader can find the raw data in a Zenodo archive, with DOI: 10.5281/zenodo.7988106.
To facilitate reproducibility of the computational environment, we execute all cases using Singularity/Apptainer images for both the PetIBM and PINN cases. 
All the container recipes are included in the repro-pack under the \texttt{resources} folder. 
The \emph{Modulus} toolkit was open-sourced by NVIDIA in March 2023,\footnote{\url{https://developer.nvidia.com/blog/physics-ml-platform-modulus-is-now-open-source/}} under the Apache License 2.0.
This is a permissive license that requires preservation of copyright and license notices and provides an express grant of patent rights. 
When we started this research, \emph{Modulus} was not yet open-source, but it was publicly available through the conditions of an End User Agreement. 
Documentation of those conditions can be found via the May 21, 2022, snapshot of the \emph{Modulus} developer website on the Internet Archive Wayback Machine.\footnote{\url{https://web.archive.org/web/20220521223413/https://catalog.ngc.nvidia.com/orgs/nvidia/teams/modulus/containers/modulus}}
We are confident that following the best practices of open science described in this statement provides good conditions for reproducibility of our results. 
Readers can inspect the code if any detail is unclear in the paper narrative, and they can re-analyze our data or re-run the computational experiments.
We spared no effort to document, organize, and preserve all the digital artifacts for this work.

    \section*{Acknowledgement}
    We appreciate the support by NVIDIA, through sponsoring the access to its high-performance computing cluster.

    \bibliography{references}

    \appendix
    \section{Supplement}

\begin{figure*} [h]
    \centering%
    \includegraphics[width=0.95\textwidth]{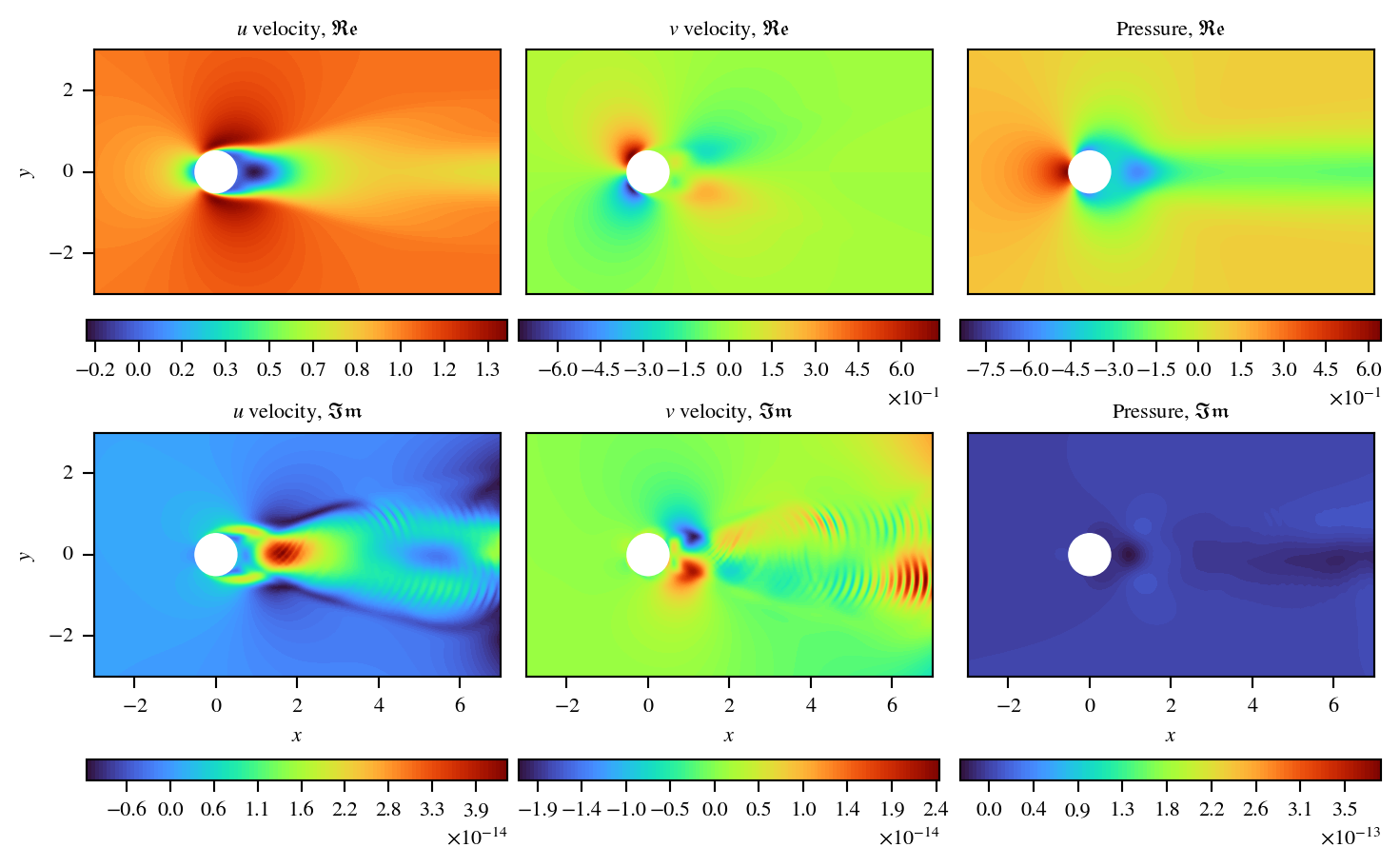}%
    \caption{%
        The \num{1}st mode in PetIBM.
    }
    \label{fig:cylinder-re200-koopman-petibm-1st}%
\end{figure*}

\begin{figure*} 
    \centering%
    \includegraphics[width=0.95\textwidth]{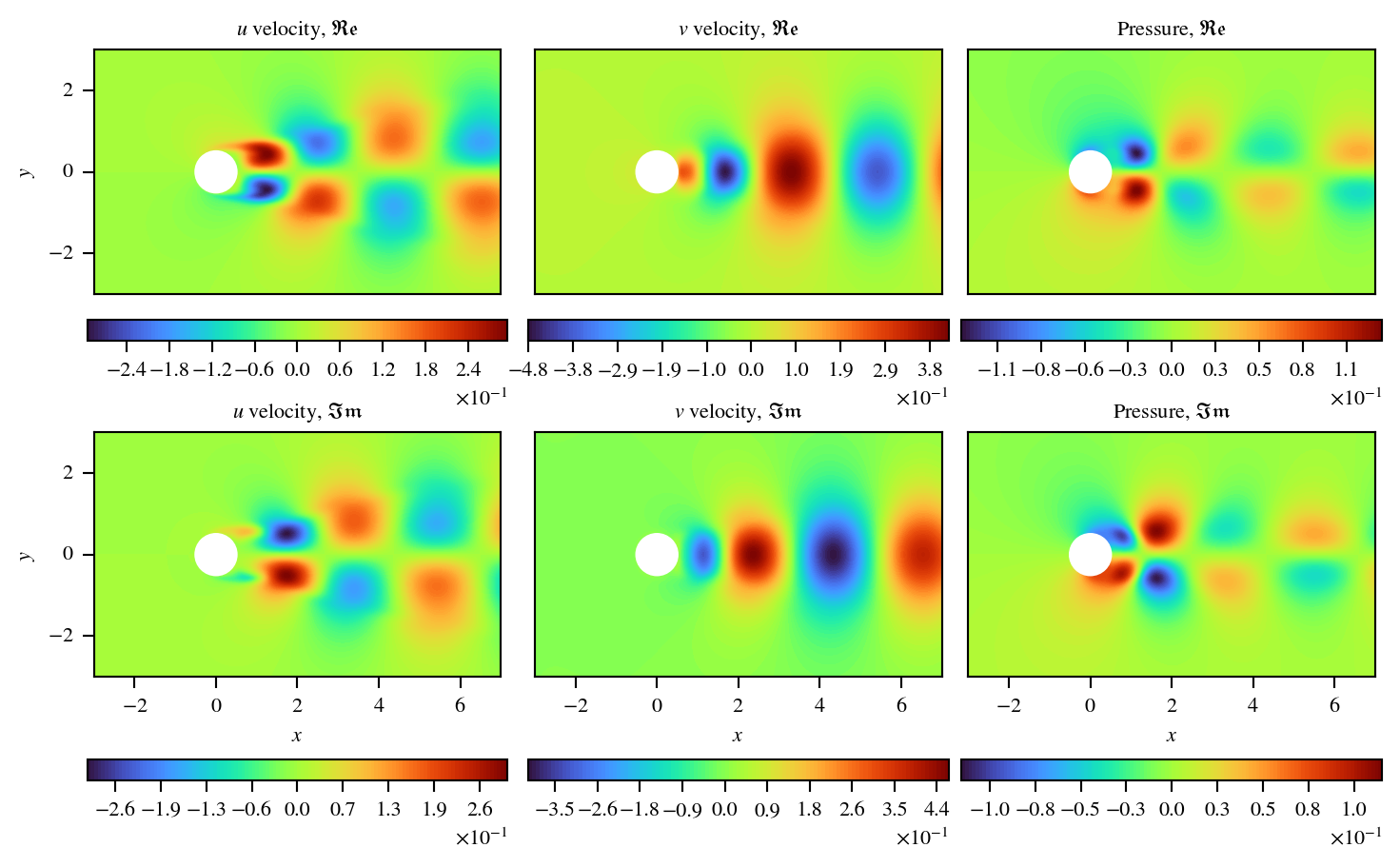}%
    \caption{%
        The \num{2}nd mode in PetIBM.
    }
    \label{fig:cylinder-re200-koopman-petibm-2nd}%
\end{figure*}

\begin{figure*} 
    \centering%
    \includegraphics[width=0.95\textwidth]{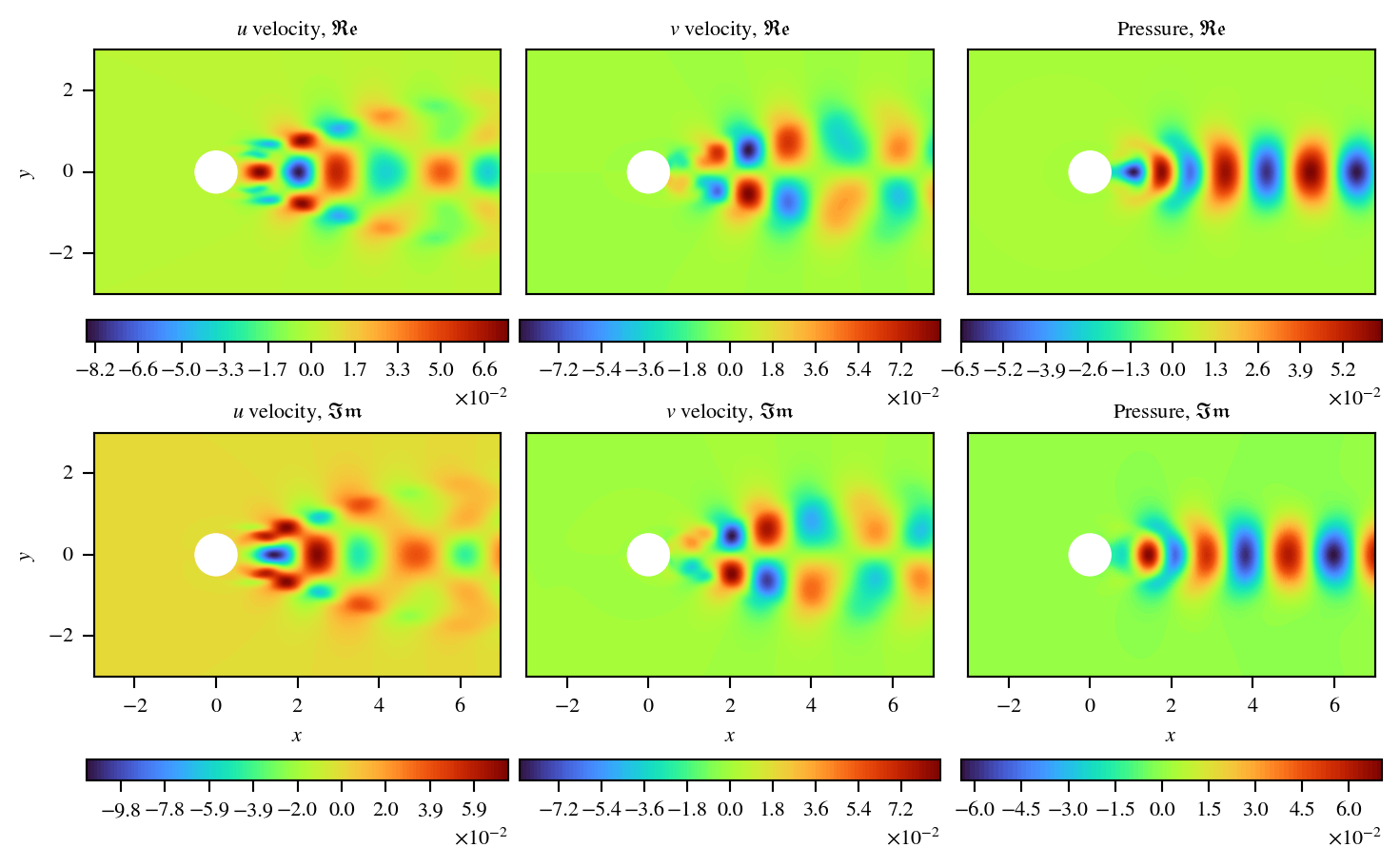}%
    \caption{%
        The \num{3}rd mode in PetIBM.
    }
    \label{fig:cylinder-re200-koopman-petibm-3rd}%
\end{figure*}

\begin{figure*} 
    \centering%
    \includegraphics[width=0.95\textwidth]{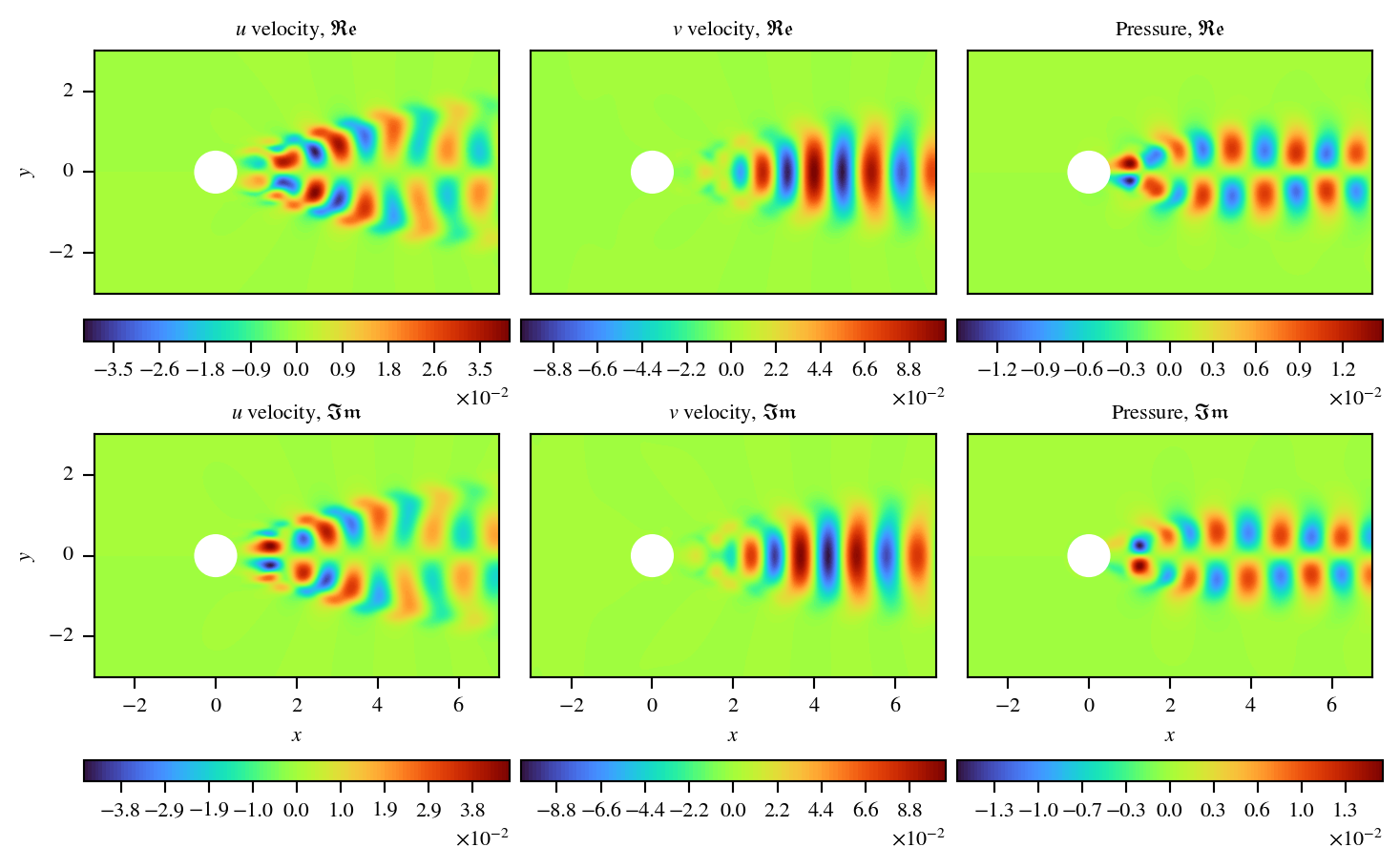}%
    \caption{%
        The \num{4}th mode in PetIBM.
    }
    \label{fig:cylinder-re200-koopman-petibm-4th}%
\end{figure*}

\begin{figure*} 
    \centering%
    \includegraphics[width=0.95\textwidth]{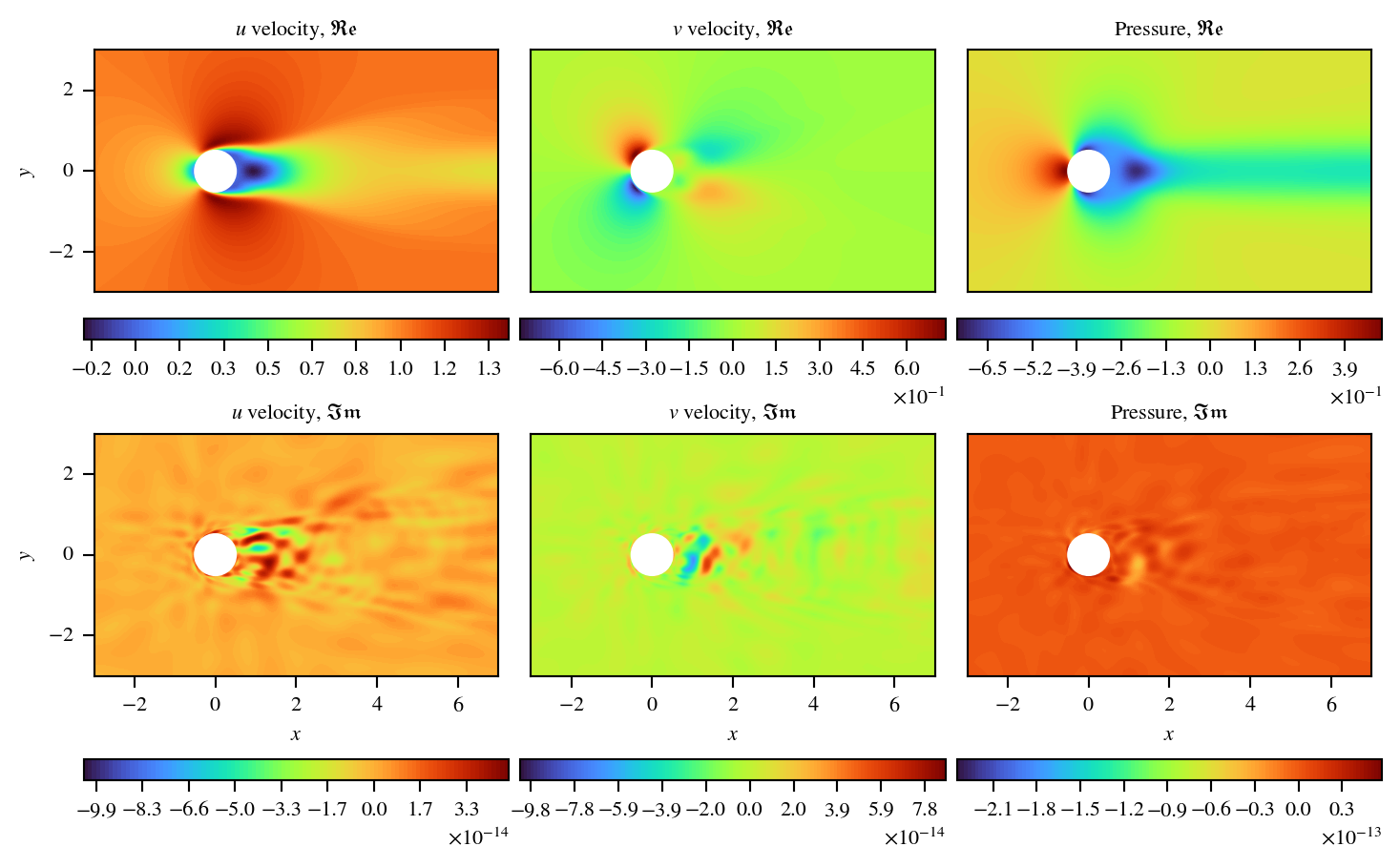}%
    \caption{%
        The \num{1}st primary mode in data-driven PINN.
    }
    \label{fig:cylinder-re200-koopman-pinn-primary-1st}%
\end{figure*}

\begin{figure*} 
    \centering%
    \includegraphics[width=0.95\textwidth]{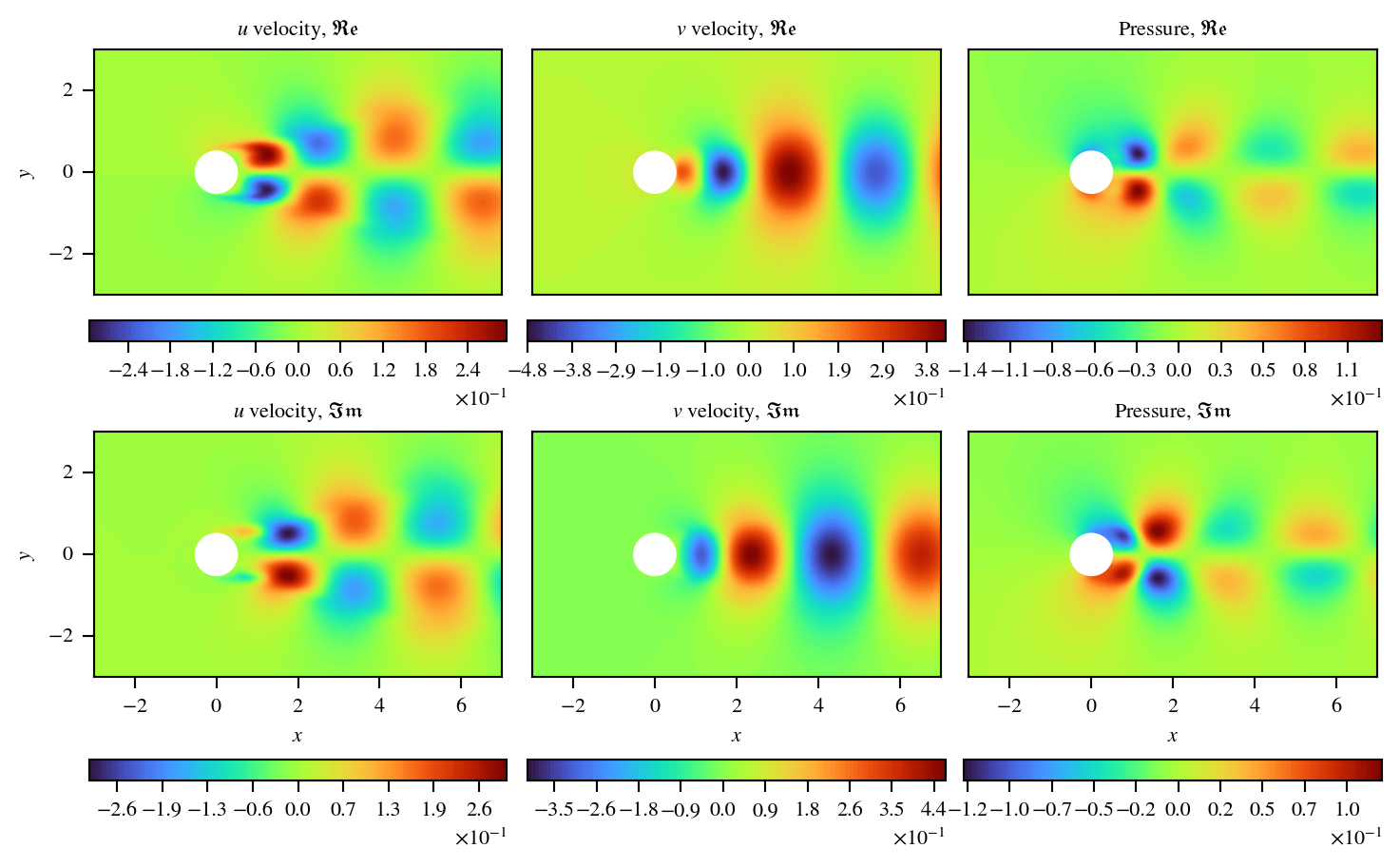}%
    \caption{%
        The \num{2}nd primary mode in data-driven PINN.
    }
    \label{fig:cylinder-re200-koopman-pinn-primary-2nd}%
\end{figure*}

\begin{figure*} 
    \centering%
    \includegraphics[width=0.95\textwidth]{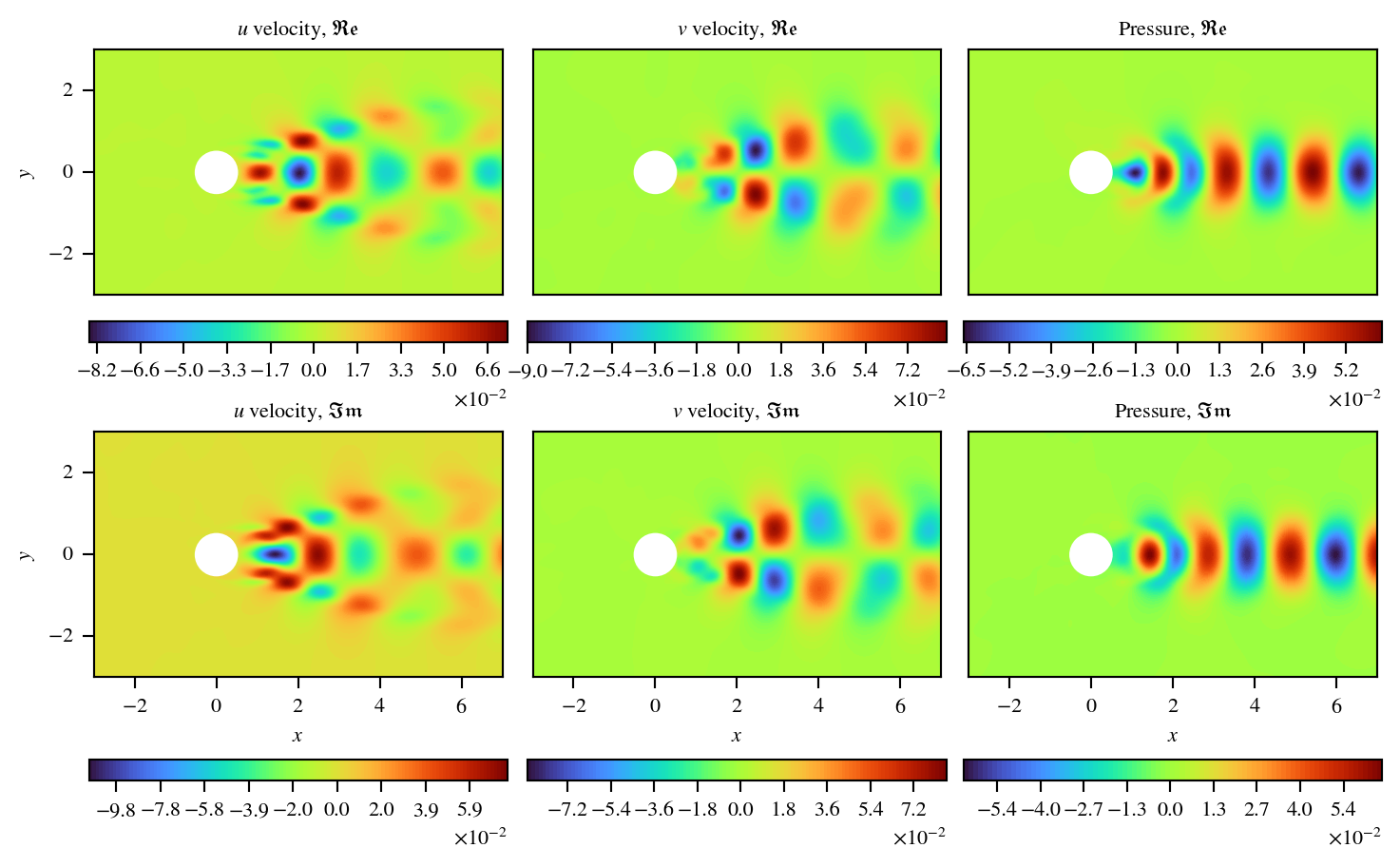}%
    \caption{%
        The \num{3}rd primary mode in data-driven PINN.
    }
    \label{fig:cylinder-re200-koopman-pinn-primary-3rd}%
\end{figure*}

\begin{figure*} 
    \centering%
    \includegraphics[width=0.95\textwidth]{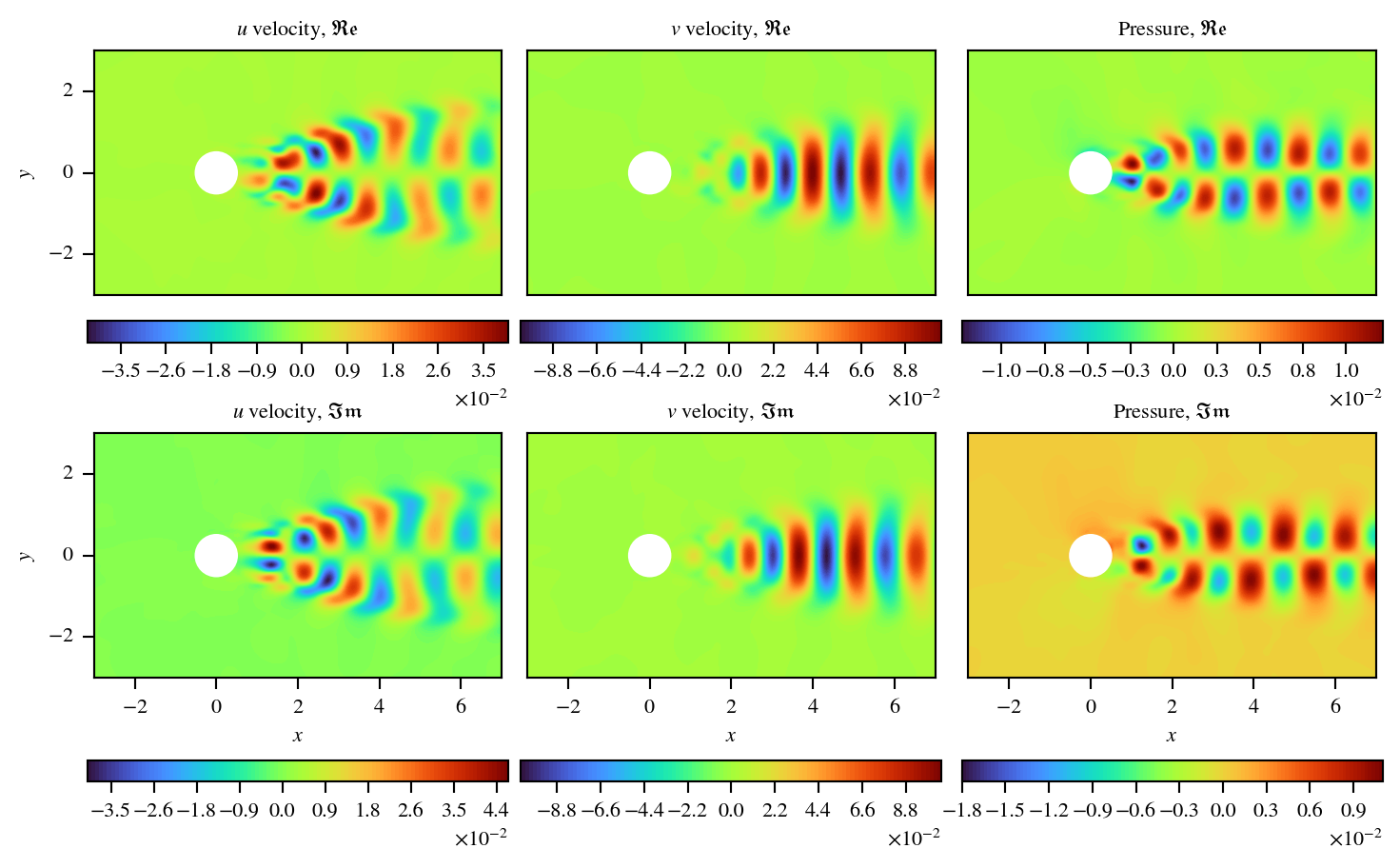}%
    \caption{%
        The \num{4}th primary mode in data-driven PINN.
    }
    \label{fig:cylinder-re200-koopman-pinn-primary-4th}%
\end{figure*}

\begin{figure*} 
    \centering%
    \includegraphics[width=0.95\textwidth]{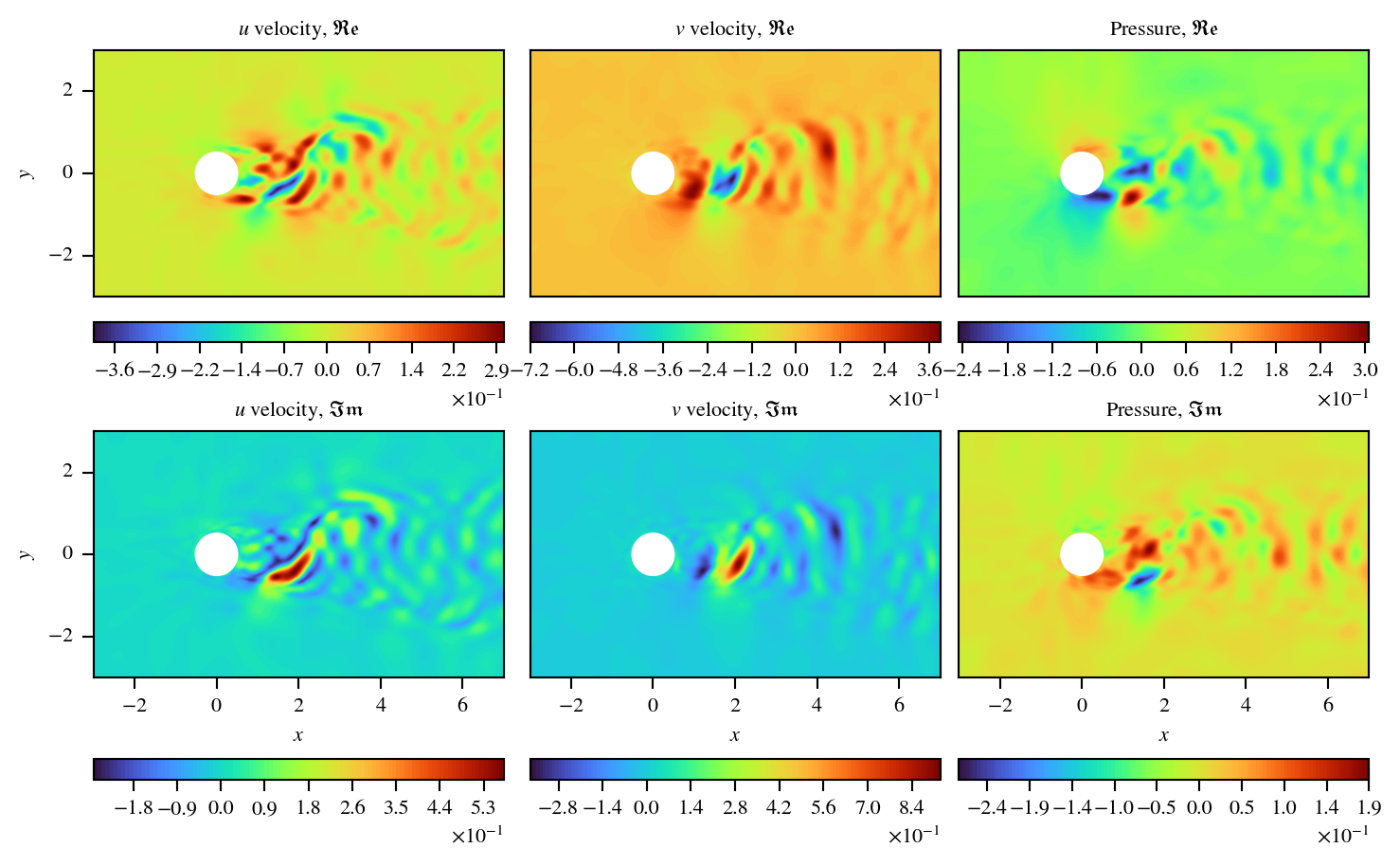}%
    \caption{%
        The \num{1}st damped mode in data-driven PINN.
    }
    \label{fig:cylinder-re200-koopman-pinn-damped-1st}%
\end{figure*}

\begin{figure*} 
    \centering%
    \includegraphics[width=0.95\textwidth]{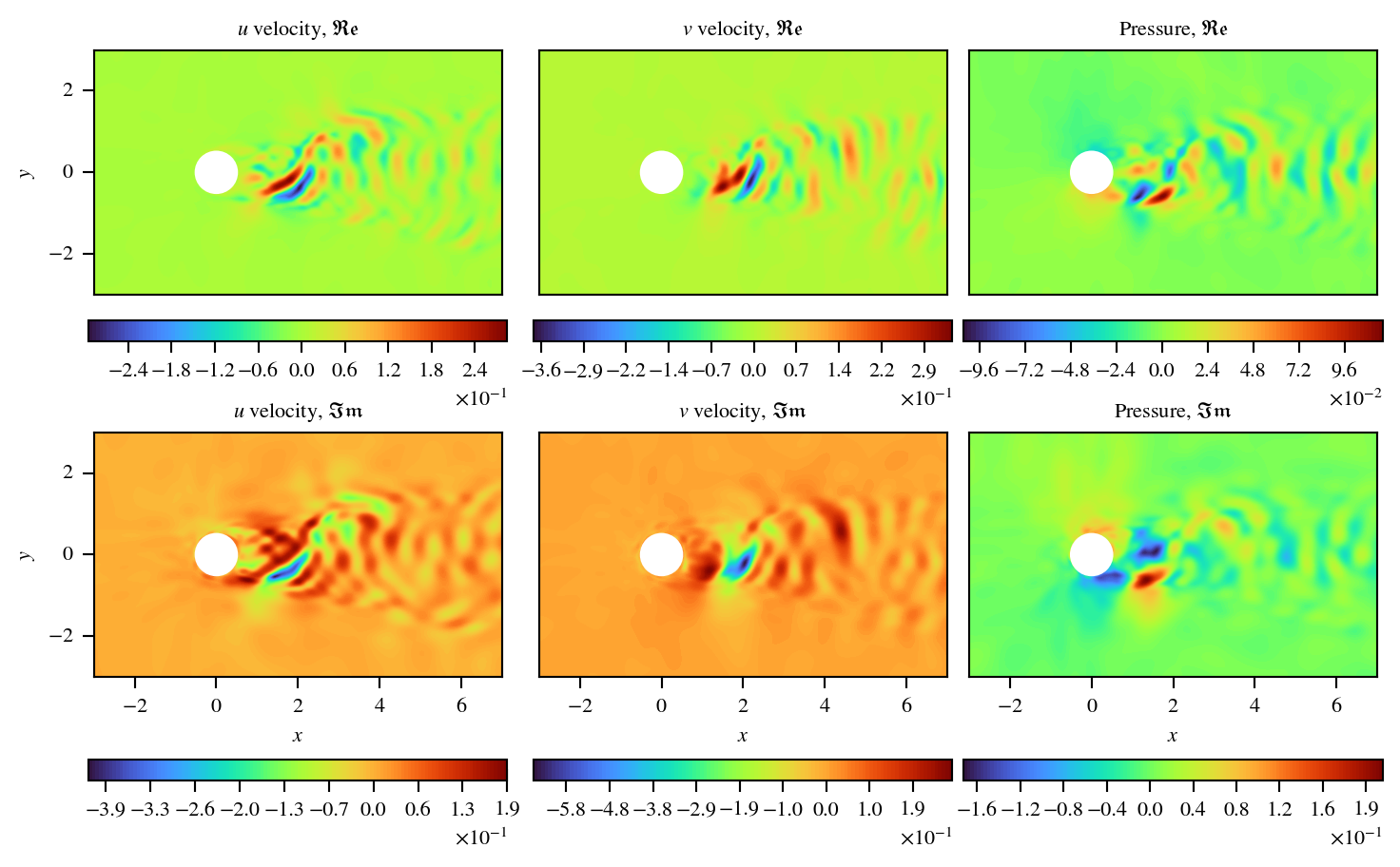}%
    \caption{%
        The \num{2}nd damped mode in data-driven PINN.
    }
    \label{fig:cylinder-re200-koopman-pinn-damped-2nd}%
\end{figure*}

\begin{figure*} 
    \centering%
    \includegraphics[width=0.95\textwidth]{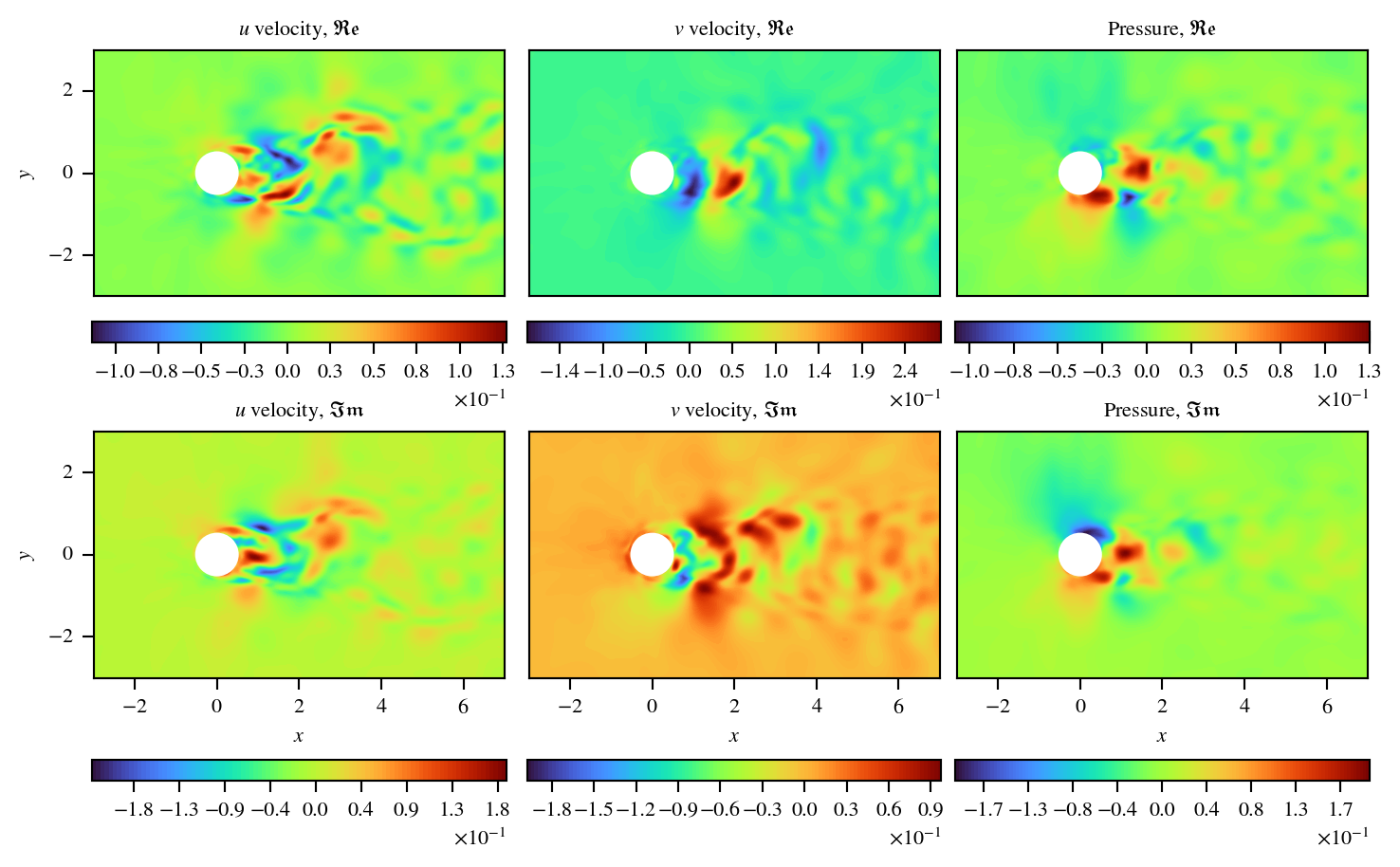}%
    \caption{%
        The \num{3}rd damped mode in data-driven PINN.
    }
    \label{fig:cylinder-re200-koopman-pinn-damped-3rd}%
\end{figure*}

\begin{figure*} 
    \centering%
    \includegraphics[width=0.95\textwidth]{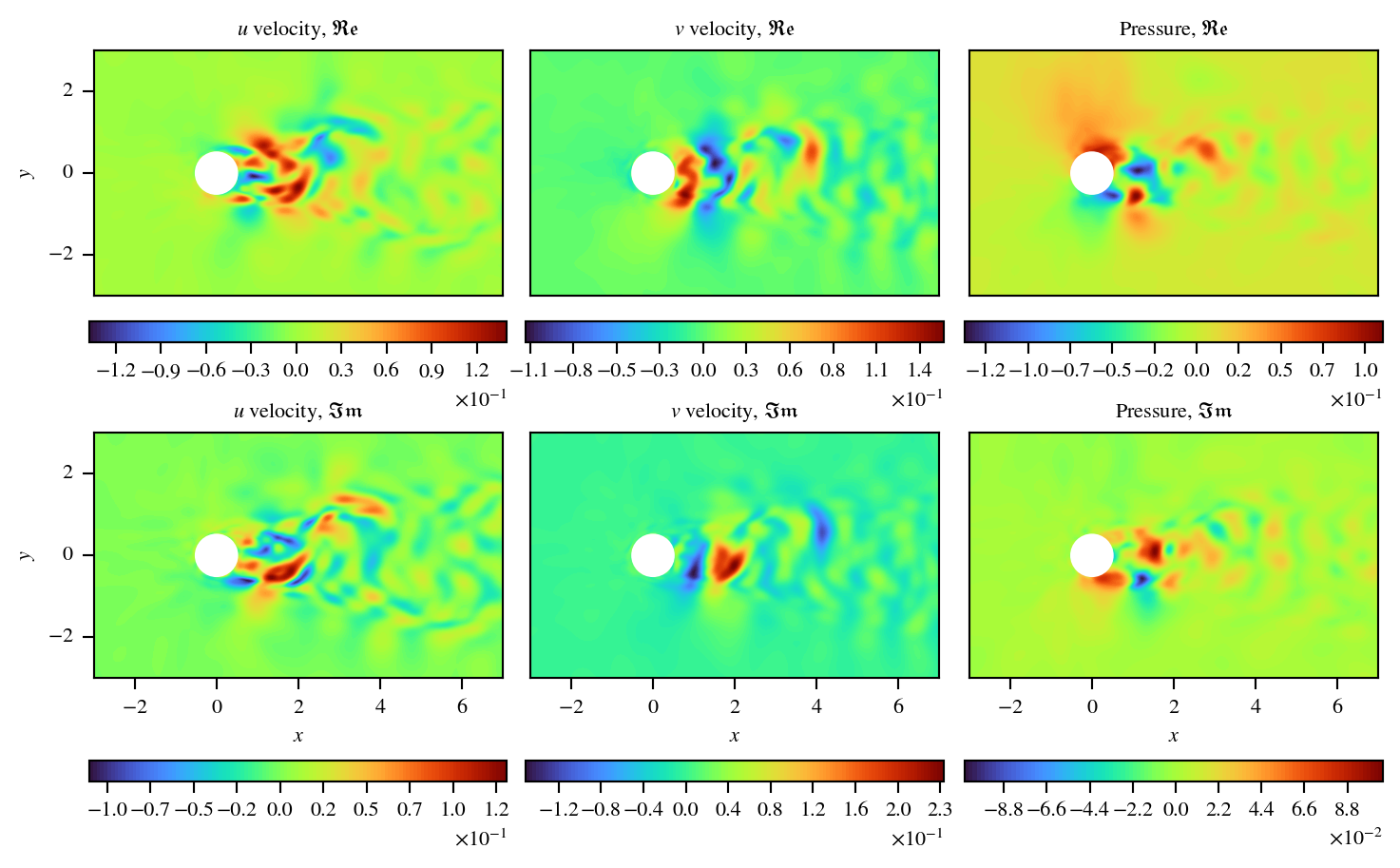}%
    \caption{%
        The \num{4}th damped mode in data-driven PINN.
    }
    \label{fig:cylinder-re200-koopman-pinn-damped-4th}%
\end{figure*}


\end{document}